\pdfoutput=1

\documentclass[aps,prx,reprint,showpacs,superscriptaddress,floatfix]{revtex4-1}
\usepackage{graphicx}
\usepackage{graphics}
\usepackage{amsmath}
\usepackage{amssymb}
\usepackage{amsfonts}
\usepackage{dcolumn}
\usepackage{dsfont}
\usepackage{latexsym}
\usepackage{rotating}
\usepackage{color}
\usepackage{latexsym}
\usepackage{bbm}
\usepackage{subfigure}
\usepackage{float}
\usepackage{epsfig}
\usepackage{epsf}
\usepackage{psfrag}
\usepackage{bm}
\usepackage{amsthm}
\usepackage{eucal}
\usepackage{mathrsfs}
\usepackage{url}
\usepackage{braket}
\usepackage{array}
\usepackage[utf8]{inputenc}
\usepackage{microtype}
\usepackage{multirow}

\usepackage{hyperref}
\hypersetup{
colorlinks=true,final=true,
        linkcolor=blue,
        citecolor=blue,
        filecolor=blue,
        urlcolor=blue,
}

\begin{document}

\title{Electronic structure and magnetic tendencies of trilayer La$_4$Ni$_3$O$_{10}$ under pressure: structural transition, molecular orbitals, and layer differentiation}

\author{Harrison LaBollita}
\email{hlabolli@asu.edu}
\affiliation{Department of Physics, Arizona State University, Tempe, AZ 85287, USA}
\author{Jesse Kapeghian}
\affiliation{Department of Physics, Arizona State University, Tempe, AZ 85287, USA}
\author{Michael R. Norman}
\affiliation{Materials Science Division, Argonne National Laboratory, Lemont, Illinois 60439, USA}
\author{Antia S. Botana}
\affiliation{Department of Physics, Arizona State University, Tempe, AZ 85287, USA}

\begin{abstract}
Motivated by the recent observation of superconductivity in the pressurized trilayer La$_4$Ni$_3$O$_{10}$ Ruddlesden-Popper (RP) nickelate, we explore its structural, electronic, and magnetic properties as a function of hydrostatic pressure from first-principles calculations. We find that in both the bilayer and trilayer nickelates, an orthorhombic(monoclinic)-to-tetragonal transition under pressure takes place concomitantly with the onset of superconductivity. The electronic structure of La$_4$Ni$_3$O$_{10}$ can be understood using a molecular trimer basis wherein $n$ molecular subbands arise as the $d_{z^2}$ orbitals hybridize strongly along the $c$-axis within the trilayer. The magnetic tendencies indicate that the ground state at ambient pressure is formed by nonmagnetic inner planes and stripe-ordered outer planes that are antiferromagnetically coupled along the $c$ axis, resulting in an unusual $\uparrow$, 0, $\downarrow$ stacking that is consistent with the spin density wave model suggested by neutron diffraction.  Such a state is destabilized by the pressures wherein superconductivity arises. Despite the presence of $d_{z^2}$ states at the Fermi level, the $d_{x^2-y^2}$ orbitals also play a key role in the electronic structure of La$_4$Ni$_3$O$_{10}$. This active role of the $d_{x^2-y^2}$ states in the low-energy physics of the trilayer RP nickelate, together with the distinct electronic behavior of inner and outer planes, resembles the physics of multilayer cuprates. 

\end{abstract}

\maketitle

\section{Introduction}
Since the discovery of high-temperature superconductivity (HTS) in the cuprates, identifying analogous materials that could help determine the essential components for HTS has posed a significant challenge in condensed matter physics \cite{Bednorz1986possible}. Recently, superconductivity was discovered in the layered nickelate family $R_{n+1}$Ni$_n$O$_{2n+2}$ ($R$ = rare-earth) providing a new perspective to this endeavor. Similar to the cuprates, these reduced layered nickelates have $n$-NiO$_2$ planes and  superconductivity in them emerges from a $3d^{9-\delta}$ ($\delta = 1/n \sim 0.2$) electron configuration on the Ni sites. The first superconducting member of the family was the infinite layer ($n=\infty$) material $R$NiO$_2$ ($R$ = La, Pr, Nd) wherein superconductivity with a maximum T$_c$$\sim$ 15 K arises upon hole doping \cite{Li2019superconductivity,Osada2020superconducting, Osada2021nickelate, Zeng2021superconductivity}. Subsequently, superconductivity was also found in the quintuple-layer ($n=5$) compound Nd$_6$Ni$_5$O$_{12}$ that exhibits a similar T$_c$ but without the need for chemical doping \cite{Pan2021superconductivity}. This series of discoveries showed that akin to the cuprates, the $R_{n+1}$Ni$_n$O$_{2n+2}$ nickelate series represents a whole family of superconductors.

The field has been recently reinvigorated after the discovery of superconductivity under pressure ($\sim$ 14 GPa) in bilayer 
La$_3$Ni$_2$O$_7$ with a maximum T$_c$ $\sim$ 80 K \cite{sun2023superconductivity,hou2023emergence,zhang2023exps}. This material with an average Ni$^{2.5+}$ ($d^{7.5}$) is the $n=2$ member of the  $R_{n+1}$Ni$_n$O$_{3n+1}$ Ruddlesden-Popper (RP) phases: the parent compounds of the reduced layered  $R_{n+1}$Ni$_n$O$_{2n+2}$ family. Unlike the reduced nickelates, the Ni atoms in these RP phases are coordinated by apical oxygens and a have nominal electronic configuration for the Ni atoms of $d^{7+\delta}$, far from that of the cuprates. The discovery of superconductivity in this material ignited an explosion of experimental \cite{filamentary,chen2023musr,wang2023i4mmmexp} and theoretical \cite{zhang2023electronic, chen2023critical, lechermann2023electronic, christiansson2023correlated, luo2023bilayer, gu2023effective, shen2023effective, zhang2023electronic, wú2023charge, yang2023possible, liu2023swave, zhang2023structural, qu2023bilayer, yang2023minimal, zhang2023trends, lu2023superconductivity, tian2023correlation,vortex2023huang, jiang2023screening, liao2023correlations, liao2023interlayer,oh2023tj,qin2023singlets, sakakibara2023hubbard, Yang2023dmrg, shilenko2023} work. Further, as with the reduced layered nickelate family, the extension of superconductivity to other members of the parent $R_{n+1}$Ni$_n$O$_{3n+1}$ RP compounds has also been perceived as a crucial problem to be tackled. Very recently, superconducting signatures in the trilayer ($n=3$) RP nickelate La$_4$Ni$_3$O$_{10}$ under pressure have also been experimentally reported with a T$_c$ $\sim$ 30 K at pressures $\sim$ 30 GPa \cite{li2023signature, zhang2023superconductivity, zhu2024superconductivity}. At ambient pressure, this material exhibits a metal-to-metal transition that results from an incommensurate density wave with both charge and magnetic character \cite{Zhang2020Intertwined}. The discovery of superconductivity under pressure in the trilayer RP nickelate unlocks the $R_{n+1}$Ni$_n$O$_{3n+1}$ series as a new family of superconducting nickelates.

\begin{figure*}
\centering
\includegraphics[width=2\columnwidth]{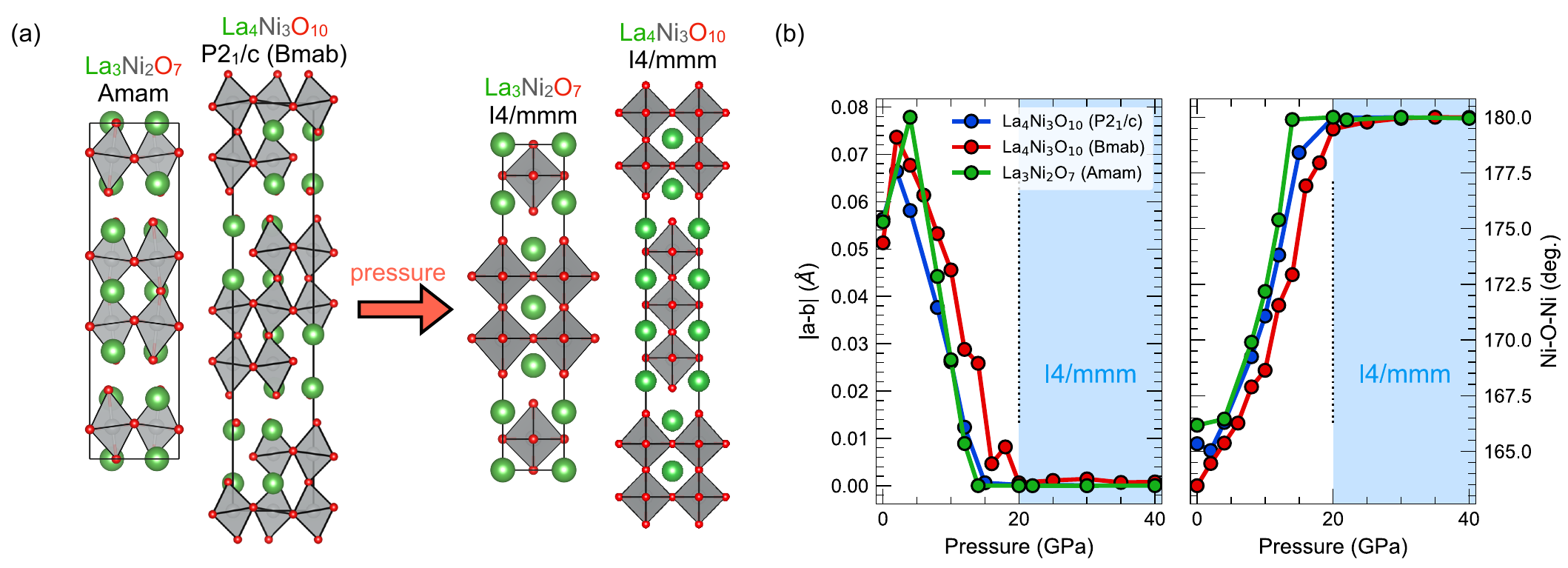}
\caption{``Tetragonalization'' of the crystal structure of La$_{n+1}$Ni$_{n}$O$_{3n+1}$ ($n=2,3$) under pressure. (a) The ambient pressure structures for La$_{3}$Ni$_{2}$O$_{7}$ ($n=2$) (orthorhombic \textit{Amam} crystal setting) and La$_{4}$Ni$_{3}$O$_{10}$ ($n=3$) (orthorhombic or monoclinc crystal settings) transform to tetragonal ($I4/mmm$) structures under pressure. Green, grey, and red spheres denote La, Ni, and O atoms, respectively. (b) Structural data capturing the structural transition with pressure: absolute difference ($|a-b|$) of the in-plane lattice constants (left) and apical Ni-O-Ni bond angle (right).}
\label{fig:4310-struct}

\end{figure*}

For an emerging novel class of superconductors, it is crucial to find the analogies and differences among the members of the family. Here, using first-principles calculations, we analyze the structural, electronic, and magnetic properties of the trilayer RP nickelate La$_4$Ni$_3$O$_{10}$ and contrast them to those of the bilayer material La$_3$Ni$_2$O$_7$.  We find that in both materials, an orthorhombic (monoclinic)-to-tetragonal transition under pressure takes place that is concomitant with the onset of superconductivity. The electronic structure of La$_4$Ni$_3$O$_{10}$ can be understood using a quantum-coupled unit -- a
molecular trimer basis -- for the trilayer wherein $n$ molecular subbands (bonding-nonbonding-antibonding) arise as the $d_{z^2}$ orbitals hybridize strongly along the $c$-axis. A similar picture has been proposed for La$_3$Ni$_2$O$_7$ \cite{zhang2023electronic}. Our spin-polarized calculations indicate that the ground state of La$_4$Ni$_3$O$_{10}$ at ambient pressure is formed by stripe-ordered magnetic outer layers (coupled antiferromagnetically along the $c$ axis), and nonmagnetic inner layers, consistent with the spin density wave model suggested by neutron diffraction data. Such a state is destabilized at the pressures in which superconductivity has been observed. Importantly, as we found for La$_3$Ni$_2$O$_7$ \cite{labollita2023electronic}, the $d_{x^2-y^2}$ orbitals play an active role in the electronic structure of the trilayer RP nickelate under pressure which, together with the distinct behavior of inner and outer layers, bears similarities with (multilayer) cuprates \cite{multilayer_cuprates}.

\section{Methods}
We have used density-functional theory (DFT)-based calculations to investigate the structural properties, electronic structure, and magnetic tendencies of La$_4$Ni$_3$O$_{10}$ under pressures up to 30 GPa~\cite{hohenberg1964inhomogeneous,kohn1965self}. La$_{4}$Ni$_{3}$O$_{10}$ crystal data in the monoclinic (orthorhombic) and tetragonal settings were taken from Refs.~\cite{Zhang2020oxygen, jung2022RPs}, respectively. Structural optimizations of the crystal structure (unit cell and internal coordinates) were performed with the plane-wave pseudopotential-based DFT code VASP~\cite{Kresse:1993bz, Kresse:1999dk, Kresse:1996kl} within the generalized gradient approximation~\cite{gga_pbe} in the nonmagnetic state. For the ambient pressure structures, only the internal coordinates were optimized. The number of plane waves in the basis was set by an energy cutoff of 520 eV. The integration in reciprocal space was carried out on an $8\times8\times2$ Monkhorst-Pack $k$-point grid. The internal forces on each atom were converged to $10^{-6}$ eV/\AA{}. Phonon calculations were performed using the frozen-phonon method as implemented in \textsc{phonopy} \cite{phonopy} interfaced with VASP for $2\times2\times1$ supercells of the monoclinic ($P2_1/c$) and tetragonal ($I4/mmm$) crystal structures.

The electronic structure for each optimized crystal structure was subsequently calculated using the all-electron, full-potential DFT code \textsc{wien2k}~\cite{Blaha2020wien2k}. The Perdew-Burke-Ernzernhof version of the generalized gradient approximation (GGA) \cite{gga_pbe} was utilized for the exchange-correlation functional.  To explore the magnetic tendencies, electronic correlation effects were included using an on-site Coulomb repulsion ($U$) for the localized Ni($3d$) states. The around mean-field (AMF) approach was employed for the double counting correction~\cite{ldau_amf} (as we showed in previous work  \cite{labollita2023electronic}, in La$_{3}$Ni$_{2}$O$_{7}$ the energetics within the fully localized limit are qualitatively similar).  We used a range of  Hubbard $U$ values from 2 to 5 eV, while the Hund's coupling $J_{\mathrm{H}}$ was fixed to the typical value of $0.7$ eV for transition-metal $3d$ electrons. An $RK_{\mathrm{max}} = 7$ was used. Muffin-tin radii (in atomic units) of 2.5, 1.95, and 1.67 for La, Ni, and O, respectively, were employed. Dense $21\times21\times21$ ($30\times29\times11$) $k$-grids for the $I4/mmm$ ($P2_{1}/c$) crystal settings were used. For the larger supercells used in some of the magnetic states we calculated, $k$-grids with similar densities were employed. For computational details on the La$_3$Ni$_2$O$_7$ calculations, we refer the reader to Ref.~\cite{labollita2023electronic}. 

\begin{figure*}
    \centering
    \includegraphics[width=1.8\columnwidth]{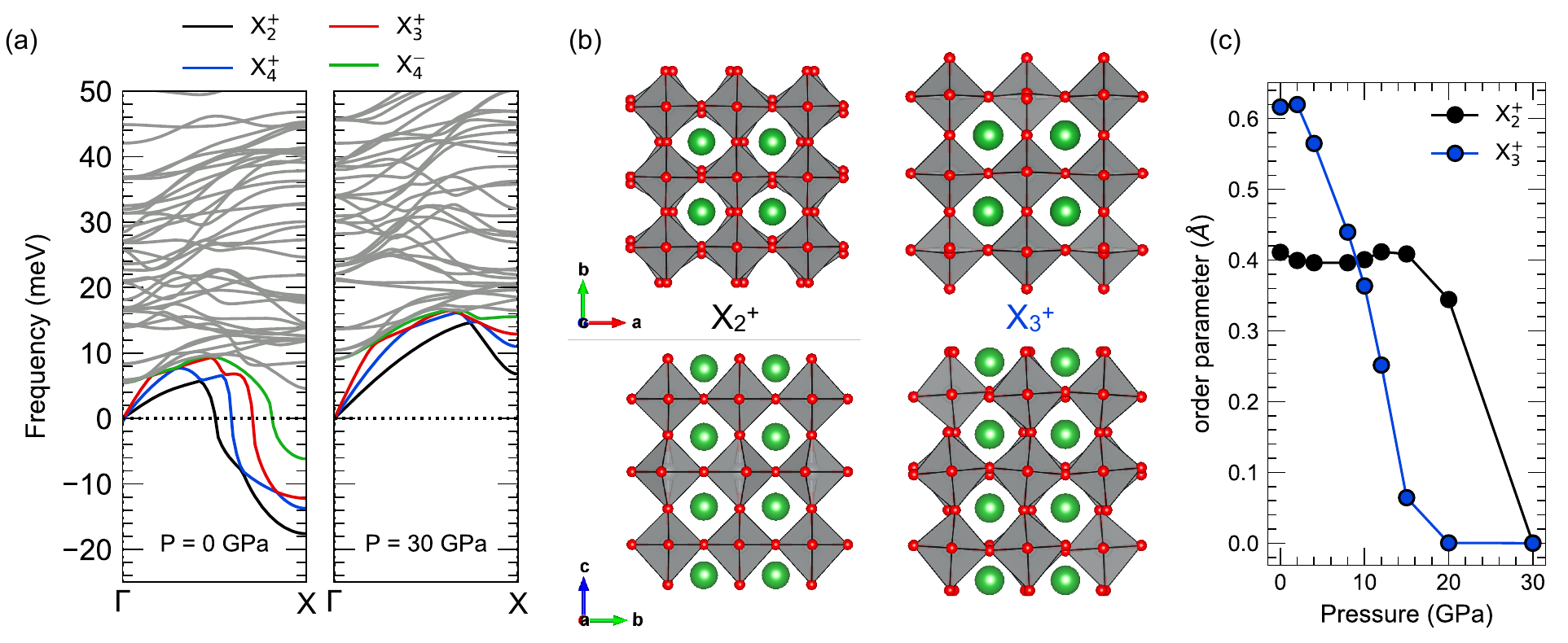}
    \caption{Lattice dynamics of La$_4$Ni$_3$O$_{10}$ under pressure. (a) Phonon dispersions at ambient (left) and 30 GPa (right) for La$_4$Ni$_3$O$_{10}$ in the tetragonal crystal setting. Colors denote phonon branches corresponding to unstable modes. (b) Crystal distortions according to the X$_{2}^{+}$ and X$_{3}^{+}$ irreducible representations (irreps) of the $I4/mmm$ space group in the $ab$ (top) and $bc$ (bottom) planes. (c) Order parameter amplitudes obtained from AMPLIMODES \cite{amplimodes1, amplimodes2} for the X$_{2}^{+}$ and X$_{3}^{+}$ irreps relative to the ambient pressure tetragonal crystal structure.}
    \label{fig:4310-phonons-main}

\end{figure*}

\section{Results}

\subsection{\label{sec:structure}`Tetragonalization' and lattice dynamics of the crystal structure under pressure}

Figure~\ref{fig:4310-struct} shows the evolution of the structural parameters of the trilayer RP La$_3$Ni$_2$O$_{10}$ with pressure (in comparison to those of bilayer La$_3$Ni$_2$O$_7$). In La$_3$Ni$_2$O$_7$, initial X-ray diffraction (XRD) experiments under pressure revealed that superconductivity is accompanied by a structural transition from a low-symmetry \textit{Amam} orthorhombic phase to a higher-symmetry (still orthorhombic) \textit{Fmmm} phase with modest pressures ($> 10-15$ GPa) \cite{sun2023superconductivity}. This structural transition is characterized by the suppression of the tilts in the NiO$_{6}$ octahedral cages. Using DFT calculations, we previously showed~\cite{labollita2023electronic} (in agreement with other theory work \cite{geisler2023structural}) that near $\sim$ 15 GPa, an orthorhombic-to-tetragonal ($I4/mmmm$) structural transition could occur instead (see Fig.~\ref{fig:4310-struct}(a,b)). Importantly, this prediction of the `tetragonalization' of the structure under pressure is in agreement with recent low temperature XRD experiments \cite{wang2023i4mmmexp}, which have confirmed that an $I4/mmm$ phase arises in La$_3$Ni$_2$O$_7$ at low temperatures ($\sim$ 40 K) when compressed to 19 GPa ~\footnote{Further ambiguity has been introduced with the report of La$_{3}$Ni$_{2}$O$_{7}$ with a `1313' stacking sequence of single and trilayer nickel planes~\cite{chen2023polymorphism, puphal2023unconventional}. Here, we only study the bilayer nickelate in the conventional (`2222') stacking and leave the comparison to the alternate stacking for future work.}.

Analogous structural relaxations for La$_4$Ni$_3$O$_{10}$ using either the monoclinic $P2_{1}/c$ or orthorhombic $Bmab$ crystal structures as a starting point \cite{Li2017Fermiology,Puggioni2018crystal, Zhang2020Intertwined,Zhang2020oxygen} indicate an identical structural transition with pressure takes place in the trilayer material (see Fig.~\ref{fig:4310-struct}(a,b)). Near $\sim 10-15$ GPa, the structure of La$_4$Ni$_3$O$_{10}$ `tetragonalizes' with the collapse of the in-plane lattice constants to the same value and the suppression of the octahedral tilts. The tetragonal structure can also be seen to become the more energetically favored structure with pressure when comparing the enthalpies of the low ($P2_{1}/c$) and high symmetry ($I4/mmm$) phases (see Fig.~\ref{fig:4310-phonons-appendix} in Appendix~\ref{app:phonons_la4310}).

To gain further insight into the lattice dynamics driving the transition towards a tetragonal structure with pressure in La$_{4}$Ni$_{3}$O$_{10}$, we calculate the evolution of the phonons in the high-symmetry tetragonal ($I4/mmm$) setting (see Fig.~\ref{fig:4310-phonons-main}). At ambient pressure we identify four unstable modes at the X point with the X$_{2}^{+}$, X$_{4}^{+}$, X$_{3}^{+}$, and X$_{4}^{-}$ irreducible representations (irreps) of the $I4/mmm$ space group (see Fig.~\ref{fig:4310-phonons-main}(a)). The X$_{2}^{+}$ irrep corresponds to the distortion of the in-plane oxygens, while the X$_{3}^{+}$, X$_{4}^{+}$, and X$_{4}^{-}$ irreps correspond to the distortions of the apical oxygens in the NiO$_6$ cage that are linked to the octahedral tilts (see Fig.~\ref{fig:4310-phonons-main}(b)). We find that these four unstable modes become quenched with pressure: Initially, the X$_{3}^{+}$ mode associated with the NiO$_{6}$ octahedral tilts is suppressed. Subsequently, the X$_{2}^{+}$ mode corresponding to the in-plane distortion of the NiO$_{6}$ octahedra is quenched (see Fig.~\ref{fig:4310-phonons-main}(c) as well as further details on the phonon calculations in Fig. \ref{fig:4310-phonons-appendix} in Appendix \ref{app:phonons_la4310}). Importantly, a symmetry mode analysis (obtained from AMPLIMODES~\cite{amplimodes1, amplimodes2}) reveals that the X$_{2}^{+}$ and X$_{3}^{+}$ distortions are the ones responsible for driving the transition from $I4/mmm$ to $P2_{1}/c$. 

Overall, our results strongly favor the transformation of the crystal structure to a tetragonal setting in both the bilayer and trilayer RP nickelates under pressure. Therefore, we conclude that the tetragonal phase in these materials is, most likely, the phase linked to the onset of superconductivity.

 \begin{figure*}
    \centering
    \includegraphics[width=2\columnwidth]{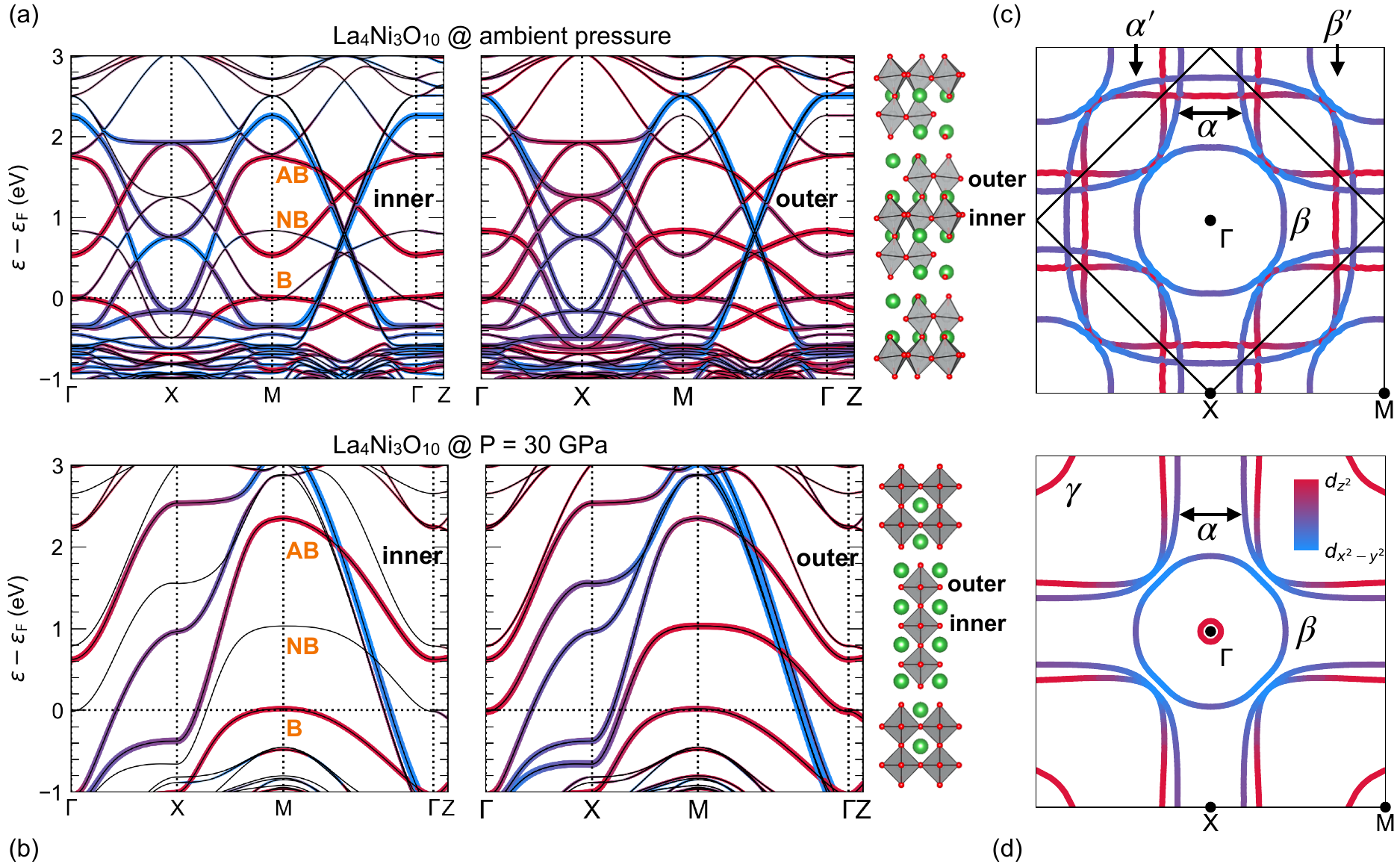}
    \caption{Electronic structure of La$_4$Ni$_3$O$_{10}$ at ambient pressure ($P2_{1}/c$, top row) and $P=30$ GPa ($I4/mmm$, bottom row). (a,b) Site- and orbital-resolved band structures along high-symmetry lines highlighting the Ni-$d_{z^{2}}$ (red) and Ni-$d_{x^{2}-y^{2}}$ (blue) orbital characters for the inner and outer inequivalent Ni sites for the ambient ($P2_{1}/c$) and 30 GPa ($I4/mmm$) phases, respectively. The Ni-$d_{z^{2}}$ bonding (B), non-bonding (NB), and anti-bonding (AB) bands are labeled. The corresponding crystal structure is shown to the right. (c,d) Corresponding Fermi surfaces for (a,b) in the $k_{z}=0$ plane with high-symmetry points and $\alpha$, $\beta$, and $\gamma$ sheets denoted ($\alpha'$ and $\beta'$ sheets are the backfolded versions of their unprimed counterparts).}
    \label{fig:4310-NM}
\end{figure*}

\subsection{Evolution of the electronic structure with pressure}
We begin our analysis by examining the pressure-induced variations in the nonmagnetic electronic structure of La$_4$Ni$_3$O$_{10}$. A formal valence count for La$_4$Ni$_3$O$_{10}$  renders an (average) Ni$^{2.67+}$ ion corresponding to a $d^{7.33}$ filling of the Ni($3d$) shell. Within each trilayer complex (three NiO$_{6}$ layers), one has (on average) four electrons in the Ni-$e_{g}$ orbitals while the Ni-$t_{2g}$ orbitals are completely filled. In order to clearly understand the band fillings in La$_4$Ni$_3$O$_{10}$, the site- and orbital-resolved band structures are shown in Fig.~\ref{fig:4310-NM}(a, b) for the ambient pressure $P2_{1}/c$ phase and for the $I4/mmm$ structure at 30 GPa, where we distinguish the Ni-$e_{g}$ ($d_{z^{2}}$ and $d_{x^{2}-y^{2}}$) orbital contributions from the inequivalent inner- and outer-layer Ni sites (note that, at ambient pressure, the electronic structure in the $Bmab$ space group is equivalent to that of the $P2_{1}/c$ phase as shown in Fig. \ref{fig:4310-YGMY} Appendix \ref{app:bands_la4310_space_group_dep}). As expected, in the vicinity of the Fermi level, both the $d_{x^2-y^2}$ and $d_{z^2}$ bands are active and strongly hybridized. It should be noted that the $d_{z^2}$ bands are filled quite differently for different Ni atoms in the structure. In contrast, no relevant distinction is observed for the filling of the $d_{x^2-y^2}$ orbitals. These differences in filling are robust with pressure and we explain their origin below.

 \textit{Molecular orbitals.--} The different band occupations for the inner and outer Ni atoms can be understood using the differences in $d_{z^2}$ filling in terms of `molecular subbands' \cite{jung2022RPs,Pardo2010quantum}. In essence, $n$ molecular subbands arise in the RP  nickelates as the $d_{z^2}$ orbitals hybridize strongly along the $c$-axis, with the molecular picture being due to the natural quantum confinement within the structure provided by the blocking slabs between successive Ni-O trilayers. As such, for the $n=3$ RP nickelate, for each Ni triple along the $c$ axis with neighbors coupled by the hopping integral $t_{\perp}$, the corresponding eigenvalues and eigenvectors are~\cite{Pardo2010quantum}:
 
\begin{equation}
\begin{aligned}
\varepsilon_{z^2} = 0,~&\pm\sqrt{2}t_{\perp}, \\
\ket{\varepsilon_{z^2}} =\frac{1}{\sqrt{2}}(1,0,-1), &~ \frac{1}{2}(1,\mp\sqrt{2},1)
\end{aligned}
\label{eqn:dimers}
\end{equation}
The odd symmetry nonbonding (NB) state ($\varepsilon=0$), which does not involve the inner Ni site, as shown by the corresponding eigenvector, is midway in energy between the even symmetry bonding (B) and antibonding (AB) states. This clean picture of molecular orbitals explains the site- and orbital-resolved band structures we obtain. Importantly, this picture survives the application of pressure and the structural phase transition, with increased splittings among the different molecular orbitals as pressure is applied. 
\begin{figure*}
    \centering
    \includegraphics[width=2\columnwidth]{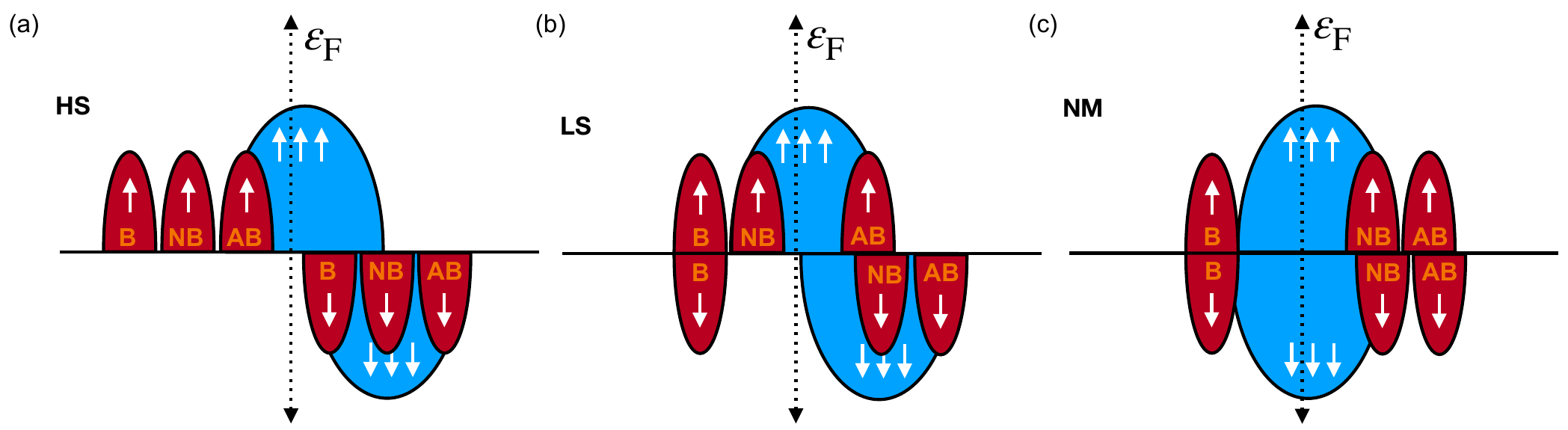}
    \caption{Level scheme showing the possible spin states for the trilayer RP nickelate La$_4$Ni$_3$O$_{10}$ using the molecular orbital picture for the Ni-$d_{z^2}$ states (three Ni atoms are represented -- within each trilayer four electrons occupy the
the Ni-e$_g$ orbitals). The $d_{x^2-y^2}$ states are depicted in blue, the $d_{z^2}$ states are depicted in red. The bonding (B), non-bonding (NB), and anti-bonding (AB) $d_{z^2}$ bands are labeled. (a) High-spin (HS). (b) Low-spin (LS). (c) Non-magnetic (NM).}
    \label{fig:spin_states}
\end{figure*}

The molecular orbital picture applies to the bilayer La$_3$Ni$_2$O$_{7}$ material as well, but in that case, one has only two molecular subbands for the dimer forming a bonding-antibonding complex (and no effective difference between the two layers exists in terms of filling, in contrast to the trilayer material). As a consequence of this splitting, in the bilayer nickelate (where we have an average Ni valence Ni$^{2.5+}$ ($d^{7.5}$), which corresponds to three electrons to be filled for each bilayer in the Ni-$e_{g}$ orbitals, with the Ni-$t_{2g}$ orbitals also being completely filled) two spin states are possible depending on how the three electrons fill the $e_g$ states: $(i)$ a low-spin (LS) state, wherein the $d_{z^2}$ bonding state is doubly occupied, whereas the antibonding state is empty, with the $d_{x^2-y^2}$ orbitals being 1/4 filled; or $(ii)$ a  high-spin (HS) state, where the bonding $d_{z^2}$ and $d_{x^2-y^2}$ orbitals are all half-filled (i.e. the majority spin channel is occupied), and the antibonding $d_{z^2}$ state remains empty. We have found \cite{labollita2023electronic} that in La$_3$Ni$_2$O$_7$ a low-spin state is favored in the high pressure/superconducting phase suggesting that the HS solution may be unfavorable for superconductivity in this material.

Returning to the trilayer nickelate case, with the above considerations about the $d_{z^2}$ splittings (and keeping in mind that the $d_{x^2-y^2}$ states are equally occupied for all the Ni atoms), how can one fill up the available four electrons in the $e_g$ levels? The different possible states are the following (see Fig.~\ref{fig:spin_states}): (\textit{i}) a nonmagnetic (NM) state, where one occupies the majority and minority bonding $d_{z^2}$ states and 1/3 of the majority and minority $d_{x^2-y^2}$ bands, (\textit{ii}) a HS state, where a $d_{z^2}$ triplet (bonding-nonbonding-antibonding) is filled as well as 1/3 of the majority $d_{x^2-y^2}$ band, and (\textit{iii}) a LS state, where the bonding $d_{z^2}$  state is doubly occupied, as well as 1/3 of the majority $d_{x^2-y^2}$ band, and the majority nonbonding $d_{z^2}$ orbital. We will return to the Ni spin states in La$_4$Ni$_3$O$_{10}$ in the next section.

 \begin{figure*}
    \centering
    \includegraphics[width=1.9\columnwidth]{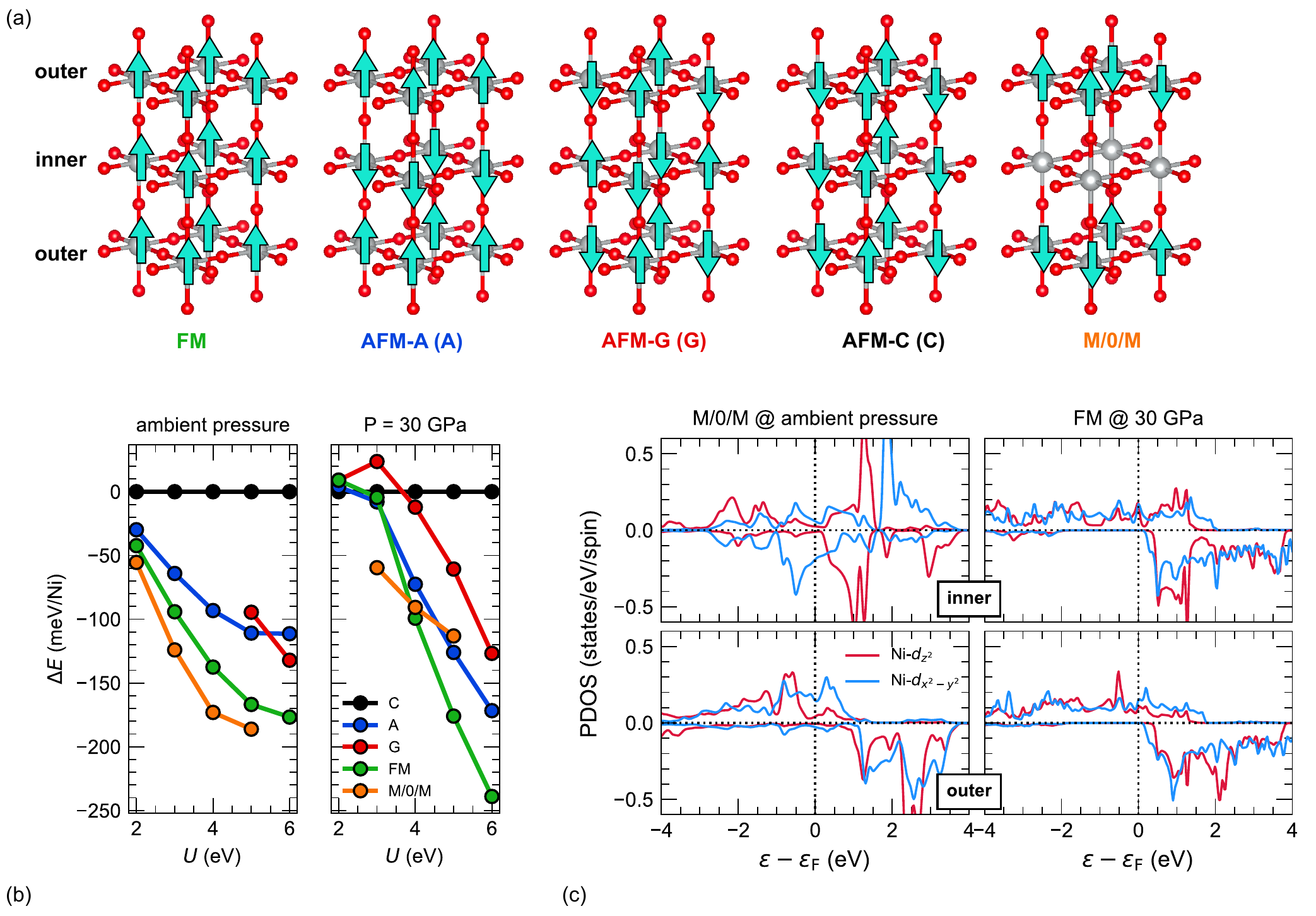}
    \caption{Magnetic tendencies in La$_4$Ni$_3$O$_{10}$ under pressure. (a) Different spin configurations within the trilayer unit attempted in our calculations. (b) Energy differences between different magnetic configurations at ambient pressure (left), for which we used the orthorhombic $Bmab$ structure, and at 30 GPa (right) for which we used our relaxed tetragonal $I4/mmm$ structure. The AFM-C (C) magnetic configuration is taken as the reference. At ambient pressure, we find that the AFM-G (G) state can only be converged for large values of the Hubbard $U$, while for smaller values it converges to the C-type solution. (c) Layer- and orbital-resolved partial density of states (PDOS) for the DFT+$U$ ground state at ambient pressure (left) and 30 GPa (right). Positive (negative) is taken for the majority (minority) spin channel. }
    \label{fig:4310-mag}
\end{figure*}

\textit{Fermi surface.--} 
To draw comparisons, we start by briefly describing the fermiology of the bilayer RP La$_{3}$Ni$_{2}$O$_{7}$ (see Fig.~\ref{fig:327-NM} in Appendix~\ref{app:327}, where we use the notation of the ARPES data for La$_4$Ni$_3$O$_{10}$ \cite{Li2017Fermiology} and refer to the different bands as $\alpha$, $\beta$, and $\gamma$). Its Fermi surface consists of one hole-like sheet of dominant $d_{x^2-y^2}$ character ($\alpha$) centered at the corner of the tetragonal zone (M),
an electron-like sheet comprised of mixed $e_g$ orbital character ($\beta$) at $\Gamma$, and another hole pocket at M of dominant $d_{z^2}$ character ($\gamma$). The nature of these Fermi surface sheets does not significantly change with pressure. We note that the importance of the $\gamma$ pocket for Cooper pairing (in $s^{\pm}$ symmetry) has been recently highlighted: if the $\gamma$ pocket vanishes, $s^{\pm}$ pairing is suppressed~\cite{zhang2023trends}.

The fermiology of La$_{4}$Ni$_{3}$O$_{10}$ is somewhat similar to that of its bilayer counterpart --with some relevant differences arising with pressure. At ambient pressure, the Fermi surface is comprised of two different types of sheets (see Fig.~\ref{fig:4310-NM}(c)): two large hole pockets centered at the corner of the tetragonal Brillouin zone ($\alpha$ band, split due to the interlayer hopping within the trilayer) and an electron pocket ($\beta$) centered at $\Gamma$. The $\alpha$ pocket has mixed $e_g$ orbital characters while the $\beta$ pocket is predominantly $d_{x^2-y^2}$ in character. The $\alpha'$ and $\beta'$ pockets are the back-folded versions of the $\alpha$ and $\beta$ pockets, respectively. We do not find the $\gamma$ pocket obtained in ARPES (with mainly Ni-$d_{z^{2}}$ character) at ambient pressure at the GGA level unless the Fermi energy is shifted (as shown in Fig. \ref{fig:4310-YGMY} in Appendix \ref{app:bands_la4310_space_group_dep}). It has been suggested~\cite{Li2017Fermiology} that the opening of a gap in the $\gamma$-$d_z^2$ band accompanies the charge density wave (CDW) formation at the metal-to-metal transition in La$_4$Ni$_3$O$_{10}$. We argue that this gap is instead due to backfolding (see Fig.~\ref{fig:4310-YGMY} in Appendix \ref{app:bands_la4310_space_group_dep} for the band structure of a $\sqrt{2}$$\times$$\sqrt{2}$ cell of the $I4/mmm$ structure, where the gap is closed).

Applying pressure to La$_{4}$Ni$_{3}$O$_{10}$, the Fermi surface exhibits many of the same features ($\alpha$ and $\beta$ pockets) with the addition of the $\gamma$ pocket of dominant $d_{z^2}$ character at the zone corner (see Fig.~\ref{fig:4310-NM}(d)). We also find a small electron pocket emerging at $\Gamma$ due to the nonbonding $d_{z^2}$ band that broadens with pressure and crosses the Fermi level. As mentioned above, the importance of the $\gamma$ pocket for Cooper pairing (in $s^{\pm}$ symmetry) has been emphasized in bilayer nickelates \cite{zhang2023trends}. With the inclusion of dynamical electronic correlations within single-site DFT+dynamical mean-field theory, the basic features of the fermiology that we describe remain the same under pressure~\cite{leonov2024electronic, tian2024effective, wang2024nonfermi}.

\subsection{Magnetic tendencies with pressure}

At ambient pressure, La$_4$Ni$_3$O$_{10}$ is known to display a metal-to-metal transition that results from intertwined charge-density and spin-density (SDW) waves~\cite{Zhang2020Intertwined}. Recent work on  La$_3$Ni$_2$O$_{7}$ also points towards the formation of density wave order \cite{chen2023musr}. Importantly, in the trilayer RP La$_4$Ni$_3$O$_{10}$, single crystal neutron diffraction data are available that were taken to test for the presence of an SDW concomitant with the CDW \cite{Zhang2020Intertwined}. The intensity distribution of the derived superlattice reflections is consistent with a rather unusual magnetic state: outer planes that are antiferromagnetically coupled and no moment on the inner planes implying a magnetic pattern of the six planes in the unit cell (three per trilayer) as follows: $\uparrow$, --, $\downarrow$; $\uparrow$, --, $\downarrow$ where -- represents a node. The slight incommensurability of the SDW ordering vector results in an approximately 5-period stripe in the plane~\cite{Zhang2020Intertwined}.

Thus to (approximately) test this magnetic state (that we denote M/0/M), as well as other types of orderings with pressure (ferromagnetic (FM), A-type antiferromagnetic (A-AFM), C-type antiferromagnetic (C-AFM) and G-type antiferromagnetic (G-AFM), as depicted in Fig.~\ref{fig:4310-mag}(a)), we performed GGA+$U$ calculations with $U$ values ranging from $2-5$ eV.  Our results of the magnetic tendencies under pressure are shown in Fig.~\ref{fig:4310-mag}(b). We find that at ambient pressure, the ground state of the system is the M/0/M state consisting of in-plane ($\pi$,0) stripe order but with a clear moment differentiation between inner and outer planes. While the inner planes have nearly quenched Ni moments  $\sim$ $0.10\mu_B$ (close to the NM state of Fig. \ref{fig:spin_states}), the outer planes consist of HS Ni atoms (with moments $\sim$ $1.2\mu_B$). Importantly, the out-of-plane coupling matches that of the superlattice reflections obtained from neutrons: $\uparrow$, --, $\downarrow$. Even though the occupations that we obtain for the different orbitals in the spin states that we quote (NM inner, HS outer) agree with the schemes of Fig. \ref{fig:spin_states}, with La$_4$Ni$_3$O$_{10}$ being an itinerant system, there are some deviations with respect to the expected orbital occupation values in the pure ionic limit (see Table \ref{tab:on-site} in Appendix \ref{app:occupations} for the orbital occupations).  At 30 GPa, the M/0/M state is destabilized in favor of a FM state for $U$ values above 4 eV, with high-spin Ni in both inner and outer planes (with Ni moments $\sim$ 1.2 $\mu_{\mathrm{B}}$). While such long-range order is unlikely at high pressure, the crucial conclusions drawn from our magnetic trends for La$_4$Ni$_3$O$_{10}$ are as follows: (1) Both ferromagnetic and antiferromagnetic couplings can be relevant. (2) The transition between magnetic ground states with pressure is driven by a redistribution in both the $d_{z^{2}}$ orbitals (that control the out-of-plane electronic structure) and the $d_{x^{2}-y^{2}}$ orbitals (that dominate the in-plane electronic structure). This suggests that the $d_{x^{2}-y^{2}}$ orbitals play a key role in the physics of La$_{4}$Ni$_{3}$O$_{10}$, as we observed in the bilayer material~\cite{labollita2023electronic}. (3) The electronic structure of the inner and outer layers in the trilayer RP nickelate exhibits distinct electronic and magnetic behavior, akin to that observed in (multilayer) cuprates \cite{multilayer_cuprates}.

\section{Summary and conclusions}
We have studied the structural, electronic, and magnetic properties of the trillyer La$_4$Ni$_3$O$_{10}$ Ruddlesden-Popper nickelate as a function of hydrostatic pressure (and compared them with its bilayer counterpart La$_3$Ni$_2$O$_7$) from a first-principles perspective. In both materials, an orthorhombic(monoclinic)-to-tetragonal transition under pressure is found, that is concomitant with the onset of superconductivity. Their electronic structure can be understood using a quantum-coupled unit, wherein $n$ molecular subbands arise as the $d_{z^2}$ orbitals overlap strongly along the $c$-axis (with a bonding-antibonding complex being formed for the bilayer and a bonding-nonbonding-antibonding complex for the trilayer) that is hybridized with active Ni-$d_{x^{2}-y^{2}}$ in the plane.  In the nonmagnetic state, the electronic structure does not change significantly with pressure, other than for the introduction of an extra pocket of $d_{z^2}$ character at the corner of the tetragonal zone in La$_4$Ni$_3$O$_{10}$.  The magnetic tendencies of the trilayer RP nickelate indicate that the ground state at ambient pressure is consistent with the spin density wave model suggested by neutron diffraction data, with distinct inner (nonmagnetic) and stripe-ordered outer Ni layers (antiferromagnetically coupled along $c$), giving rise to an overall $\uparrow$, 0, $\downarrow$ stacking. Such a state is destabilized by the pressures wherein superconductivity arises. The distinct behavior of the inner and outer planes in the trilayer RP nickelate indicates that superconductivity might arise from layer differentiation, similar to multilayer cuprates.

\section*{Acknowledgements}
We acknowledge NSF Grant No. DMR-2045826 and the ASU Research Computing Center for HPC resources.
MRN was supported by the Materials Sciences and Engineering Division, Basic Energy Sciences, Office of Science, U.S. Dept. of Energy.

\bibliography{references.bib}

\begin{thebibliography}{67}%
\makeatletter
\providecommand \@ifxundefined [1]{%
 \@ifx{#1\undefined}
}%
\providecommand \@ifnum [1]{%
 \ifnum #1\expandafter \@firstoftwo
 \else \expandafter \@secondoftwo
 \fi
}%
\providecommand \@ifx [1]{%
 \ifx #1\expandafter \@firstoftwo
 \else \expandafter \@secondoftwo
 \fi
}%
\providecommand \natexlab [1]{#1}%
\providecommand \enquote  [1]{``#1''}%
\providecommand \bibnamefont  [1]{#1}%
\providecommand \bibfnamefont [1]{#1}%
\providecommand \citenamefont [1]{#1}%
\providecommand \href@noop [0]{\@secondoftwo}%
\providecommand \href [0]{\begingroup \@sanitize@url \@href}%
\providecommand \@href[1]{\@@startlink{#1}\@@href}%
\providecommand \@@href[1]{\endgroup#1\@@endlink}%
\providecommand \@sanitize@url [0]{\catcode `\\12\catcode `\$12\catcode `\&12\catcode `\#12\catcode `\^12\catcode `\_12\catcode `\%12\relax}%
\providecommand \@@startlink[1]{}%
\providecommand \@@endlink[0]{}%
\providecommand \url  [0]{\begingroup\@sanitize@url \@url }%
\providecommand \@url [1]{\endgroup\@href {#1}{\urlprefix }}%
\providecommand \urlprefix  [0]{URL }%
\providecommand \Eprint [0]{\href }%
\providecommand \doibase [0]{http://dx.doi.org/}%
\providecommand \selectlanguage [0]{\@gobble}%
\providecommand \bibinfo  [0]{\@secondoftwo}%
\providecommand \bibfield  [0]{\@secondoftwo}%
\providecommand \translation [1]{[#1]}%
\providecommand \BibitemOpen [0]{}%
\providecommand \bibitemStop [0]{}%
\providecommand \bibitemNoStop [0]{.\EOS\space}%
\providecommand \EOS [0]{\spacefactor3000\relax}%
\providecommand \BibitemShut  [1]{\csname bibitem#1\endcsname}%
\let\auto@bib@innerbib\@empty
\bibitem [{\citenamefont {Bednorz}\ and\ \citenamefont {M\"{u}ller}(1986)}]{Bednorz1986possible}%
  \BibitemOpen
  \bibfield  {author} {\bibinfo {author} {\bibfnamefont {J.}~\bibnamefont {Bednorz}}\ and\ \bibinfo {author} {\bibfnamefont {K.}~\bibnamefont {M\"{u}ller}},\ }\href {\doibase https://doi.org/10.1007/BF01303701} {\bibfield  {journal} {\bibinfo  {journal} {Z. Phys. B}\ }\textbf {\bibinfo {volume} {64}},\ \bibinfo {pages} {189} (\bibinfo {year} {1986})}\BibitemShut {NoStop}%
\bibitem [{\citenamefont {Li}\ \emph {et~al.}(2019)\citenamefont {Li}, \citenamefont {Lee}, \citenamefont {Wang}, \citenamefont {Osada}, \citenamefont {Crossley}, \citenamefont {Lee}, \citenamefont {Cui}, \citenamefont {Hikita},\ and\ \citenamefont {Hwang}}]{Li2019superconductivity}%
  \BibitemOpen
  \bibfield  {author} {\bibinfo {author} {\bibfnamefont {D.}~\bibnamefont {Li}}, \bibinfo {author} {\bibfnamefont {K.}~\bibnamefont {Lee}}, \bibinfo {author} {\bibfnamefont {B.~Y.}\ \bibnamefont {Wang}}, \bibinfo {author} {\bibfnamefont {M.}~\bibnamefont {Osada}}, \bibinfo {author} {\bibfnamefont {S.}~\bibnamefont {Crossley}}, \bibinfo {author} {\bibfnamefont {H.~R.}\ \bibnamefont {Lee}}, \bibinfo {author} {\bibfnamefont {Y.}~\bibnamefont {Cui}}, \bibinfo {author} {\bibfnamefont {Y.}~\bibnamefont {Hikita}}, \ and\ \bibinfo {author} {\bibfnamefont {H.~Y.}\ \bibnamefont {Hwang}},\ }\href {\doibase 10.1038/s41586-019-1496-5} {\bibfield  {journal} {\bibinfo  {journal} {Nature}\ }\textbf {\bibinfo {volume} {572}},\ \bibinfo {pages} {624} (\bibinfo {year} {2019})}\BibitemShut {NoStop}%
\bibitem [{\citenamefont {Osada}\ \emph {et~al.}(2020)\citenamefont {Osada}, \citenamefont {Wang}, \citenamefont {Goodge}, \citenamefont {Lee}, \citenamefont {Yoon}, \citenamefont {Sakuma}, \citenamefont {Li}, \citenamefont {Miura}, \citenamefont {Kourkoutis},\ and\ \citenamefont {Hwang}}]{Osada2020superconducting}%
  \BibitemOpen
  \bibfield  {author} {\bibinfo {author} {\bibfnamefont {M.}~\bibnamefont {Osada}}, \bibinfo {author} {\bibfnamefont {B.~Y.}\ \bibnamefont {Wang}}, \bibinfo {author} {\bibfnamefont {B.~H.}\ \bibnamefont {Goodge}}, \bibinfo {author} {\bibfnamefont {K.}~\bibnamefont {Lee}}, \bibinfo {author} {\bibfnamefont {H.}~\bibnamefont {Yoon}}, \bibinfo {author} {\bibfnamefont {K.}~\bibnamefont {Sakuma}}, \bibinfo {author} {\bibfnamefont {D.}~\bibnamefont {Li}}, \bibinfo {author} {\bibfnamefont {M.}~\bibnamefont {Miura}}, \bibinfo {author} {\bibfnamefont {L.~F.}\ \bibnamefont {Kourkoutis}}, \ and\ \bibinfo {author} {\bibfnamefont {H.~Y.}\ \bibnamefont {Hwang}},\ }\href {\doibase 10.1021/acs.nanolett.0c01392} {\bibfield  {journal} {\bibinfo  {journal} {Nano Letters}\ }\textbf {\bibinfo {volume} {20}},\ \bibinfo {pages} {5735} (\bibinfo {year} {2020})}\BibitemShut {NoStop}%
\bibitem [{\citenamefont {Osada}\ \emph {et~al.}(2021)\citenamefont {Osada}, \citenamefont {Wang}, \citenamefont {Goodge}, \citenamefont {Harvey}, \citenamefont {Lee}, \citenamefont {Li}, \citenamefont {Kourkoutis},\ and\ \citenamefont {Hwang}}]{Osada2021nickelate}%
  \BibitemOpen
  \bibfield  {author} {\bibinfo {author} {\bibfnamefont {M.}~\bibnamefont {Osada}}, \bibinfo {author} {\bibfnamefont {B.~Y.}\ \bibnamefont {Wang}}, \bibinfo {author} {\bibfnamefont {B.~H.}\ \bibnamefont {Goodge}}, \bibinfo {author} {\bibfnamefont {S.~P.}\ \bibnamefont {Harvey}}, \bibinfo {author} {\bibfnamefont {K.}~\bibnamefont {Lee}}, \bibinfo {author} {\bibfnamefont {D.}~\bibnamefont {Li}}, \bibinfo {author} {\bibfnamefont {L.~F.}\ \bibnamefont {Kourkoutis}}, \ and\ \bibinfo {author} {\bibfnamefont {H.~Y.}\ \bibnamefont {Hwang}},\ }\href {\doibase 10.1002/adma.202104083} {\bibfield  {journal} {\bibinfo  {journal} {Adv. Mater.}\ }\textbf {\bibinfo {volume} {33}},\ \bibinfo {pages} {2104083} (\bibinfo {year} {2021})}\BibitemShut {NoStop}%
\bibitem [{\citenamefont {Zeng}\ \emph {et~al.}(2022)\citenamefont {Zeng}, \citenamefont {Li}, \citenamefont {Chow}, \citenamefont {Cao}, \citenamefont {Zhang}, \citenamefont {Tang}, \citenamefont {Yin}, \citenamefont {Lim}, \citenamefont {Hu}, \citenamefont {Yang},\ and\ \citenamefont {Ariando}}]{Zeng2021superconductivity}%
  \BibitemOpen
  \bibfield  {author} {\bibinfo {author} {\bibfnamefont {S.}~\bibnamefont {Zeng}}, \bibinfo {author} {\bibfnamefont {C.}~\bibnamefont {Li}}, \bibinfo {author} {\bibfnamefont {L.~E.}\ \bibnamefont {Chow}}, \bibinfo {author} {\bibfnamefont {Y.}~\bibnamefont {Cao}}, \bibinfo {author} {\bibfnamefont {Z.}~\bibnamefont {Zhang}}, \bibinfo {author} {\bibfnamefont {C.~S.}\ \bibnamefont {Tang}}, \bibinfo {author} {\bibfnamefont {X.}~\bibnamefont {Yin}}, \bibinfo {author} {\bibfnamefont {Z.~S.}\ \bibnamefont {Lim}}, \bibinfo {author} {\bibfnamefont {J.}~\bibnamefont {Hu}}, \bibinfo {author} {\bibfnamefont {P.}~\bibnamefont {Yang}}, \ and\ \bibinfo {author} {\bibfnamefont {A.}~\bibnamefont {Ariando}},\ }\href {\doibase 10.1126/sciadv.abl9927} {\bibfield  {journal} {\bibinfo  {journal} {Sci Adv}\ }\textbf {\bibinfo {volume} {8}},\ \bibinfo {pages} {eabl9927} (\bibinfo {year} {2022})}\BibitemShut {NoStop}%
\bibitem [{\citenamefont {Pan}\ \emph {et~al.}(2022)\citenamefont {Pan}, \citenamefont {Ferenc~Segedin}, \citenamefont {LaBollita}, \citenamefont {Song}, \citenamefont {Nica}, \citenamefont {Goodge}, \citenamefont {Pierce}, \citenamefont {Doyle}, \citenamefont {Novakov}, \citenamefont {C{\'o}rdova~Carrizales}, \citenamefont {N'Diaye}, \citenamefont {Shafer}, \citenamefont {Paik}, \citenamefont {Heron}, \citenamefont {Mason}, \citenamefont {Yacoby}, \citenamefont {Kourkoutis}, \citenamefont {Erten}, \citenamefont {Brooks}, \citenamefont {Botana},\ and\ \citenamefont {Mundy}}]{Pan2021superconductivity}%
  \BibitemOpen
  \bibfield  {author} {\bibinfo {author} {\bibfnamefont {G.~A.}\ \bibnamefont {Pan}}, \bibinfo {author} {\bibfnamefont {D.}~\bibnamefont {Ferenc~Segedin}}, \bibinfo {author} {\bibfnamefont {H.}~\bibnamefont {LaBollita}}, \bibinfo {author} {\bibfnamefont {Q.}~\bibnamefont {Song}}, \bibinfo {author} {\bibfnamefont {E.~M.}\ \bibnamefont {Nica}}, \bibinfo {author} {\bibfnamefont {B.~H.}\ \bibnamefont {Goodge}}, \bibinfo {author} {\bibfnamefont {A.~T.}\ \bibnamefont {Pierce}}, \bibinfo {author} {\bibfnamefont {S.}~\bibnamefont {Doyle}}, \bibinfo {author} {\bibfnamefont {S.}~\bibnamefont {Novakov}}, \bibinfo {author} {\bibfnamefont {D.}~\bibnamefont {C{\'o}rdova~Carrizales}}, \bibinfo {author} {\bibfnamefont {A.~T.}\ \bibnamefont {N'Diaye}}, \bibinfo {author} {\bibfnamefont {P.}~\bibnamefont {Shafer}}, \bibinfo {author} {\bibfnamefont {H.}~\bibnamefont {Paik}}, \bibinfo {author} {\bibfnamefont {J.~T.}\ \bibnamefont {Heron}}, \bibinfo {author} {\bibfnamefont {J.~A.}\ \bibnamefont {Mason}}, \bibinfo {author}
  {\bibfnamefont {A.}~\bibnamefont {Yacoby}}, \bibinfo {author} {\bibfnamefont {L.~F.}\ \bibnamefont {Kourkoutis}}, \bibinfo {author} {\bibfnamefont {O.}~\bibnamefont {Erten}}, \bibinfo {author} {\bibfnamefont {C.~M.}\ \bibnamefont {Brooks}}, \bibinfo {author} {\bibfnamefont {A.~S.}\ \bibnamefont {Botana}}, \ and\ \bibinfo {author} {\bibfnamefont {J.~A.}\ \bibnamefont {Mundy}},\ }\href {\doibase 10.1038/s41563-021-01142-9} {\bibfield  {journal} {\bibinfo  {journal} {Nat. Mater.}\ }\textbf {\bibinfo {volume} {21}},\ \bibinfo {pages} {160} (\bibinfo {year} {2022})}\BibitemShut {NoStop}%
\bibitem [{\citenamefont {Sun}\ \emph {et~al.}(2023)\citenamefont {Sun}, \citenamefont {Huo}, \citenamefont {Hu}, \citenamefont {Li}, \citenamefont {Liu}, \citenamefont {Han}, \citenamefont {Tang}, \citenamefont {Mao}, \citenamefont {Yang}, \citenamefont {Wang}, \citenamefont {Cheng}, \citenamefont {Yao}, \citenamefont {Zhang},\ and\ \citenamefont {Wang}}]{sun2023superconductivity}%
  \BibitemOpen
  \bibfield  {author} {\bibinfo {author} {\bibfnamefont {H.}~\bibnamefont {Sun}}, \bibinfo {author} {\bibfnamefont {M.}~\bibnamefont {Huo}}, \bibinfo {author} {\bibfnamefont {X.}~\bibnamefont {Hu}}, \bibinfo {author} {\bibfnamefont {J.}~\bibnamefont {Li}}, \bibinfo {author} {\bibfnamefont {Z.}~\bibnamefont {Liu}}, \bibinfo {author} {\bibfnamefont {Y.}~\bibnamefont {Han}}, \bibinfo {author} {\bibfnamefont {L.}~\bibnamefont {Tang}}, \bibinfo {author} {\bibfnamefont {Z.}~\bibnamefont {Mao}}, \bibinfo {author} {\bibfnamefont {P.}~\bibnamefont {Yang}}, \bibinfo {author} {\bibfnamefont {B.}~\bibnamefont {Wang}}, \bibinfo {author} {\bibfnamefont {J.}~\bibnamefont {Cheng}}, \bibinfo {author} {\bibfnamefont {D.-X.}\ \bibnamefont {Yao}}, \bibinfo {author} {\bibfnamefont {G.-M.}\ \bibnamefont {Zhang}}, \ and\ \bibinfo {author} {\bibfnamefont {M.}~\bibnamefont {Wang}},\ }\href {\doibase 10.1038/s41586-023-06408-7} {\bibfield  {journal} {\bibinfo  {journal} {Nature}\ }\textbf {\bibinfo {volume} {621}},\ \bibinfo {pages}
  {493} (\bibinfo {year} {2023})}\BibitemShut {NoStop}%
\bibitem [{\citenamefont {Hou}\ \emph {et~al.}(2023)\citenamefont {Hou}, \citenamefont {Yang}, \citenamefont {Liu}, \citenamefont {Li}, \citenamefont {Shan}, \citenamefont {Ma}, \citenamefont {Wang}, \citenamefont {Wang}, \citenamefont {Guo}, \citenamefont {Sun}, \citenamefont {Uwatoko}, \citenamefont {Wang}, \citenamefont {Zhang}, \citenamefont {Wang},\ and\ \citenamefont {Cheng}}]{hou2023emergence}%
  \BibitemOpen
  \bibfield  {author} {\bibinfo {author} {\bibfnamefont {J.}~\bibnamefont {Hou}}, \bibinfo {author} {\bibfnamefont {P.-T.}\ \bibnamefont {Yang}}, \bibinfo {author} {\bibfnamefont {Z.-Y.}\ \bibnamefont {Liu}}, \bibinfo {author} {\bibfnamefont {J.-Y.}\ \bibnamefont {Li}}, \bibinfo {author} {\bibfnamefont {P.-F.}\ \bibnamefont {Shan}}, \bibinfo {author} {\bibfnamefont {L.}~\bibnamefont {Ma}}, \bibinfo {author} {\bibfnamefont {G.}~\bibnamefont {Wang}}, \bibinfo {author} {\bibfnamefont {N.-N.}\ \bibnamefont {Wang}}, \bibinfo {author} {\bibfnamefont {H.-Z.}\ \bibnamefont {Guo}}, \bibinfo {author} {\bibfnamefont {J.-P.}\ \bibnamefont {Sun}}, \bibinfo {author} {\bibfnamefont {Y.}~\bibnamefont {Uwatoko}}, \bibinfo {author} {\bibfnamefont {M.}~\bibnamefont {Wang}}, \bibinfo {author} {\bibfnamefont {G.-M.}\ \bibnamefont {Zhang}}, \bibinfo {author} {\bibfnamefont {B.-S.}\ \bibnamefont {Wang}}, \ and\ \bibinfo {author} {\bibfnamefont {J.-G.}\ \bibnamefont {Cheng}},\ }\href {\doibase 10.1088/0256-307X/40/11/117302}
  {\bibfield  {journal} {\bibinfo  {journal} {Chin. Phys. Lett.}\ }\textbf {\bibinfo {volume} {40}},\ \bibinfo {eid} {117302} (\bibinfo {year} {2023})}\BibitemShut {NoStop}%
\bibitem [{\citenamefont {Zhang}\ \emph {et~al.}(2023{\natexlab{a}})\citenamefont {Zhang}, \citenamefont {Su}, \citenamefont {Huang}, \citenamefont {Sun}, \citenamefont {Huo}, \citenamefont {Shan}, \citenamefont {Ye}, \citenamefont {Yang}, \citenamefont {Li}, \citenamefont {Smidman}, \citenamefont {Wang}, \citenamefont {Jiao},\ and\ \citenamefont {Yuan}}]{zhang2023exps}%
  \BibitemOpen
  \bibfield  {author} {\bibinfo {author} {\bibfnamefont {Y.}~\bibnamefont {Zhang}}, \bibinfo {author} {\bibfnamefont {D.}~\bibnamefont {Su}}, \bibinfo {author} {\bibfnamefont {Y.}~\bibnamefont {Huang}}, \bibinfo {author} {\bibfnamefont {H.}~\bibnamefont {Sun}}, \bibinfo {author} {\bibfnamefont {M.}~\bibnamefont {Huo}}, \bibinfo {author} {\bibfnamefont {Z.}~\bibnamefont {Shan}}, \bibinfo {author} {\bibfnamefont {K.}~\bibnamefont {Ye}}, \bibinfo {author} {\bibfnamefont {Z.}~\bibnamefont {Yang}}, \bibinfo {author} {\bibfnamefont {R.}~\bibnamefont {Li}}, \bibinfo {author} {\bibfnamefont {M.}~\bibnamefont {Smidman}}, \bibinfo {author} {\bibfnamefont {M.}~\bibnamefont {Wang}}, \bibinfo {author} {\bibfnamefont {L.}~\bibnamefont {Jiao}}, \ and\ \bibinfo {author} {\bibfnamefont {H.}~\bibnamefont {Yuan}},\ }\href {https://arxiv.org/abs/2307.14819} {\bibfield  {journal} {\bibinfo  {journal} {arXiv:2307.14819}\ } (\bibinfo {year} {2023}{\natexlab{a}})}\BibitemShut {NoStop}%
\bibitem [{\citenamefont {Zhou}\ \emph {et~al.}(2023)\citenamefont {Zhou}, \citenamefont {Guo}, \citenamefont {Cai}, \citenamefont {Sun}, \citenamefont {Wang}, \citenamefont {Zhao}, \citenamefont {Han}, \citenamefont {Chen}, \citenamefont {Wu}, \citenamefont {Ding}, \citenamefont {Wang}, \citenamefont {Xiang}, \citenamefont {kwang Mao},\ and\ \citenamefont {Sun}}]{filamentary}%
  \BibitemOpen
  \bibfield  {author} {\bibinfo {author} {\bibfnamefont {Y.}~\bibnamefont {Zhou}}, \bibinfo {author} {\bibfnamefont {J.}~\bibnamefont {Guo}}, \bibinfo {author} {\bibfnamefont {S.}~\bibnamefont {Cai}}, \bibinfo {author} {\bibfnamefont {H.}~\bibnamefont {Sun}}, \bibinfo {author} {\bibfnamefont {P.}~\bibnamefont {Wang}}, \bibinfo {author} {\bibfnamefont {J.}~\bibnamefont {Zhao}}, \bibinfo {author} {\bibfnamefont {J.}~\bibnamefont {Han}}, \bibinfo {author} {\bibfnamefont {X.}~\bibnamefont {Chen}}, \bibinfo {author} {\bibfnamefont {Q.}~\bibnamefont {Wu}}, \bibinfo {author} {\bibfnamefont {Y.}~\bibnamefont {Ding}}, \bibinfo {author} {\bibfnamefont {M.}~\bibnamefont {Wang}}, \bibinfo {author} {\bibfnamefont {T.}~\bibnamefont {Xiang}}, \bibinfo {author} {\bibfnamefont {H.}~\bibnamefont {kwang Mao}}, \ and\ \bibinfo {author} {\bibfnamefont {L.}~\bibnamefont {Sun}},\ }\href {https://arxiv.org/abs/2311.12361} {\bibfield  {journal} {\bibinfo  {journal} {arXiv:2311.12361}\ } (\bibinfo {year} {2023})}\BibitemShut {NoStop}%
\bibitem [{\citenamefont {Chen}\ \emph {et~al.}(2023{\natexlab{a}})\citenamefont {Chen}, \citenamefont {Liu}, \citenamefont {Jiao}, \citenamefont {Zou}, \citenamefont {Luo}, \citenamefont {Wu}, \citenamefont {Zhang}, \citenamefont {Guo},\ and\ \citenamefont {Shu}}]{chen2023musr}%
  \BibitemOpen
  \bibfield  {author} {\bibinfo {author} {\bibfnamefont {K.}~\bibnamefont {Chen}}, \bibinfo {author} {\bibfnamefont {X.}~\bibnamefont {Liu}}, \bibinfo {author} {\bibfnamefont {J.}~\bibnamefont {Jiao}}, \bibinfo {author} {\bibfnamefont {M.}~\bibnamefont {Zou}}, \bibinfo {author} {\bibfnamefont {Y.}~\bibnamefont {Luo}}, \bibinfo {author} {\bibfnamefont {Q.}~\bibnamefont {Wu}}, \bibinfo {author} {\bibfnamefont {N.}~\bibnamefont {Zhang}}, \bibinfo {author} {\bibfnamefont {Y.}~\bibnamefont {Guo}}, \ and\ \bibinfo {author} {\bibfnamefont {L.}~\bibnamefont {Shu}},\ }\href {https://arxiv.org/abs/2311.15717} {\bibfield  {journal} {\bibinfo  {journal} {arXiv:2311.15717}\ } (\bibinfo {year} {2023}{\natexlab{a}})}\BibitemShut {NoStop}%
\bibitem [{\citenamefont {Wang}\ \emph {et~al.}(2023)\citenamefont {Wang}, \citenamefont {Li}, \citenamefont {Xie}, \citenamefont {Liu}, \citenamefont {Sun}, \citenamefont {Huang}, \citenamefont {Gao}, \citenamefont {Nakagawa}, \citenamefont {Fu}, \citenamefont {Dong}, \citenamefont {Cao}, \citenamefont {Yu}, \citenamefont {Kawaguchi}, \citenamefont {Kadobayashi}, \citenamefont {Wang}, \citenamefont {Jin}, \citenamefont {Mao},\ and\ \citenamefont {Liu}}]{wang2023i4mmmexp}%
  \BibitemOpen
  \bibfield  {author} {\bibinfo {author} {\bibfnamefont {L.}~\bibnamefont {Wang}}, \bibinfo {author} {\bibfnamefont {Y.}~\bibnamefont {Li}}, \bibinfo {author} {\bibfnamefont {S.}~\bibnamefont {Xie}}, \bibinfo {author} {\bibfnamefont {F.}~\bibnamefont {Liu}}, \bibinfo {author} {\bibfnamefont {H.}~\bibnamefont {Sun}}, \bibinfo {author} {\bibfnamefont {C.}~\bibnamefont {Huang}}, \bibinfo {author} {\bibfnamefont {Y.}~\bibnamefont {Gao}}, \bibinfo {author} {\bibfnamefont {T.}~\bibnamefont {Nakagawa}}, \bibinfo {author} {\bibfnamefont {B.}~\bibnamefont {Fu}}, \bibinfo {author} {\bibfnamefont {B.}~\bibnamefont {Dong}}, \bibinfo {author} {\bibfnamefont {Z.}~\bibnamefont {Cao}}, \bibinfo {author} {\bibfnamefont {R.}~\bibnamefont {Yu}}, \bibinfo {author} {\bibfnamefont {S.~I.}\ \bibnamefont {Kawaguchi}}, \bibinfo {author} {\bibfnamefont {H.}~\bibnamefont {Kadobayashi}}, \bibinfo {author} {\bibfnamefont {M.}~\bibnamefont {Wang}}, \bibinfo {author} {\bibfnamefont {C.}~\bibnamefont {Jin}}, \bibinfo {author} {\bibfnamefont
  {H.-K.}\ \bibnamefont {Mao}}, \ and\ \bibinfo {author} {\bibfnamefont {H.}~\bibnamefont {Liu}},\ }\href {https://arxiv.org/abs/2311.09186} {\bibfield  {journal} {\bibinfo  {journal} {arXiv:2311.09186}\ } (\bibinfo {year} {2023})}\BibitemShut {NoStop}%
\bibitem [{\citenamefont {Zhang}\ \emph {et~al.}(2023{\natexlab{b}})\citenamefont {Zhang}, \citenamefont {Lin}, \citenamefont {Moreo},\ and\ \citenamefont {Dagotto}}]{zhang2023electronic}%
  \BibitemOpen
  \bibfield  {author} {\bibinfo {author} {\bibfnamefont {Y.}~\bibnamefont {Zhang}}, \bibinfo {author} {\bibfnamefont {L.-F.}\ \bibnamefont {Lin}}, \bibinfo {author} {\bibfnamefont {A.}~\bibnamefont {Moreo}}, \ and\ \bibinfo {author} {\bibfnamefont {E.}~\bibnamefont {Dagotto}},\ }\href {\doibase 10.1103/PhysRevB.108.L180510} {\bibfield  {journal} {\bibinfo  {journal} {Phys. Rev. B}\ }\textbf {\bibinfo {volume} {108}},\ \bibinfo {pages} {L180510} (\bibinfo {year} {2023}{\natexlab{b}})}\BibitemShut {NoStop}%
\bibitem [{\citenamefont {Chen}\ \emph {et~al.}(2023{\natexlab{b}})\citenamefont {Chen}, \citenamefont {Jiang}, \citenamefont {Li}, \citenamefont {Zhong},\ and\ \citenamefont {Lu}}]{chen2023critical}%
  \BibitemOpen
  \bibfield  {author} {\bibinfo {author} {\bibfnamefont {X.}~\bibnamefont {Chen}}, \bibinfo {author} {\bibfnamefont {P.}~\bibnamefont {Jiang}}, \bibinfo {author} {\bibfnamefont {J.}~\bibnamefont {Li}}, \bibinfo {author} {\bibfnamefont {Z.}~\bibnamefont {Zhong}}, \ and\ \bibinfo {author} {\bibfnamefont {Y.}~\bibnamefont {Lu}},\ }\href {https://arxiv.org/abs/2307.07154} {\bibfield  {journal} {\bibinfo  {journal} {arXiv:2307.07154}\ } (\bibinfo {year} {2023}{\natexlab{b}})}\BibitemShut {NoStop}%
\bibitem [{\citenamefont {Lechermann}\ \emph {et~al.}(2023)\citenamefont {Lechermann}, \citenamefont {Gondolf}, \citenamefont {B\"otzel},\ and\ \citenamefont {Eremin}}]{lechermann2023electronic}%
  \BibitemOpen
  \bibfield  {author} {\bibinfo {author} {\bibfnamefont {F.}~\bibnamefont {Lechermann}}, \bibinfo {author} {\bibfnamefont {J.}~\bibnamefont {Gondolf}}, \bibinfo {author} {\bibfnamefont {S.}~\bibnamefont {B\"otzel}}, \ and\ \bibinfo {author} {\bibfnamefont {I.~M.}\ \bibnamefont {Eremin}},\ }\href {\doibase 10.1103/PhysRevB.108.L201121} {\bibfield  {journal} {\bibinfo  {journal} {Phys. Rev. B}\ }\textbf {\bibinfo {volume} {108}},\ \bibinfo {pages} {L201121} (\bibinfo {year} {2023})}\BibitemShut {NoStop}%
\bibitem [{\citenamefont {Christiansson}\ \emph {et~al.}(2023)\citenamefont {Christiansson}, \citenamefont {Petocchi},\ and\ \citenamefont {Werner}}]{christiansson2023correlated}%
  \BibitemOpen
  \bibfield  {author} {\bibinfo {author} {\bibfnamefont {V.}~\bibnamefont {Christiansson}}, \bibinfo {author} {\bibfnamefont {F.}~\bibnamefont {Petocchi}}, \ and\ \bibinfo {author} {\bibfnamefont {P.}~\bibnamefont {Werner}},\ }\href {\doibase 10.1103/PhysRevLett.131.206501} {\bibfield  {journal} {\bibinfo  {journal} {Phys. Rev. Lett.}\ }\textbf {\bibinfo {volume} {131}},\ \bibinfo {pages} {206501} (\bibinfo {year} {2023})}\BibitemShut {NoStop}%
\bibitem [{\citenamefont {Luo}\ \emph {et~al.}(2023)\citenamefont {Luo}, \citenamefont {Hu}, \citenamefont {Wang}, \citenamefont {W\'u},\ and\ \citenamefont {Yao}}]{luo2023bilayer}%
  \BibitemOpen
  \bibfield  {author} {\bibinfo {author} {\bibfnamefont {Z.}~\bibnamefont {Luo}}, \bibinfo {author} {\bibfnamefont {X.}~\bibnamefont {Hu}}, \bibinfo {author} {\bibfnamefont {M.}~\bibnamefont {Wang}}, \bibinfo {author} {\bibfnamefont {W.}~\bibnamefont {W\'u}}, \ and\ \bibinfo {author} {\bibfnamefont {D.-X.}\ \bibnamefont {Yao}},\ }\href {\doibase 10.1103/PhysRevLett.131.126001} {\bibfield  {journal} {\bibinfo  {journal} {Phys. Rev. Lett.}\ }\textbf {\bibinfo {volume} {131}},\ \bibinfo {pages} {126001} (\bibinfo {year} {2023})}\BibitemShut {NoStop}%
\bibitem [{\citenamefont {Gu}\ \emph {et~al.}(2023)\citenamefont {Gu}, \citenamefont {Le}, \citenamefont {Yang}, \citenamefont {Wu},\ and\ \citenamefont {Hu}}]{gu2023effective}%
  \BibitemOpen
  \bibfield  {author} {\bibinfo {author} {\bibfnamefont {Y.}~\bibnamefont {Gu}}, \bibinfo {author} {\bibfnamefont {C.}~\bibnamefont {Le}}, \bibinfo {author} {\bibfnamefont {Z.}~\bibnamefont {Yang}}, \bibinfo {author} {\bibfnamefont {X.}~\bibnamefont {Wu}}, \ and\ \bibinfo {author} {\bibfnamefont {J.}~\bibnamefont {Hu}},\ }\href {https://arxiv.org/abs/2306.07275} {\bibfield  {journal} {\bibinfo  {journal} {arXiv:2306.07275}\ } (\bibinfo {year} {2023})}\BibitemShut {NoStop}%
\bibitem [{\citenamefont {Shen}\ \emph {et~al.}(2023{\natexlab{a}})\citenamefont {Shen}, \citenamefont {Qin},\ and\ \citenamefont {Zhang}}]{shen2023effective}%
  \BibitemOpen
  \bibfield  {author} {\bibinfo {author} {\bibfnamefont {Y.}~\bibnamefont {Shen}}, \bibinfo {author} {\bibfnamefont {M.}~\bibnamefont {Qin}}, \ and\ \bibinfo {author} {\bibfnamefont {G.-M.}\ \bibnamefont {Zhang}},\ }\href {\doibase 10.1088/0256-307X/40/12/127401} {\bibfield  {journal} {\bibinfo  {journal} {Chin. Phys. Lett.}\ }\textbf {\bibinfo {volume} {40}},\ \bibinfo {pages} {127401} (\bibinfo {year} {2023}{\natexlab{a}})}\BibitemShut {NoStop}%
\bibitem [{\citenamefont {Wú}\ \emph {et~al.}(2023)\citenamefont {Wú}, \citenamefont {Luo}, \citenamefont {Yao},\ and\ \citenamefont {Wang}}]{wú2023charge}%
  \BibitemOpen
  \bibfield  {author} {\bibinfo {author} {\bibfnamefont {W.}~\bibnamefont {Wú}}, \bibinfo {author} {\bibfnamefont {Z.}~\bibnamefont {Luo}}, \bibinfo {author} {\bibfnamefont {D.-X.}\ \bibnamefont {Yao}}, \ and\ \bibinfo {author} {\bibfnamefont {M.}~\bibnamefont {Wang}},\ }\href {https://arxiv.org/abs/2307.05662} {\bibfield  {journal} {\bibinfo  {journal} {arXiv:2307.05662}\ } (\bibinfo {year} {2023})}\BibitemShut {NoStop}%
\bibitem [{\citenamefont {Yang}\ \emph {et~al.}(2023{\natexlab{a}})\citenamefont {Yang}, \citenamefont {Wang},\ and\ \citenamefont {Wang}}]{yang2023possible}%
  \BibitemOpen
  \bibfield  {author} {\bibinfo {author} {\bibfnamefont {Q.-G.}\ \bibnamefont {Yang}}, \bibinfo {author} {\bibfnamefont {D.}~\bibnamefont {Wang}}, \ and\ \bibinfo {author} {\bibfnamefont {Q.-H.}\ \bibnamefont {Wang}},\ }\href {\doibase 10.1103/PhysRevB.108.L140505} {\bibfield  {journal} {\bibinfo  {journal} {Phys. Rev. B}\ }\textbf {\bibinfo {volume} {108}},\ \bibinfo {pages} {L140505} (\bibinfo {year} {2023}{\natexlab{a}})}\BibitemShut {NoStop}%
\bibitem [{\citenamefont {Liu}\ \emph {et~al.}(2023)\citenamefont {Liu}, \citenamefont {Mei}, \citenamefont {Ye}, \citenamefont {Chen},\ and\ \citenamefont {Yang}}]{liu2023swave}%
  \BibitemOpen
  \bibfield  {author} {\bibinfo {author} {\bibfnamefont {Y.-B.}\ \bibnamefont {Liu}}, \bibinfo {author} {\bibfnamefont {J.-W.}\ \bibnamefont {Mei}}, \bibinfo {author} {\bibfnamefont {F.}~\bibnamefont {Ye}}, \bibinfo {author} {\bibfnamefont {W.-Q.}\ \bibnamefont {Chen}}, \ and\ \bibinfo {author} {\bibfnamefont {F.}~\bibnamefont {Yang}},\ }\href {\doibase 10.1103/PhysRevLett.131.236002} {\bibfield  {journal} {\bibinfo  {journal} {Phys. Rev. Lett.}\ }\textbf {\bibinfo {volume} {131}},\ \bibinfo {pages} {236002} (\bibinfo {year} {2023})}\BibitemShut {NoStop}%
\bibitem [{\citenamefont {Zhang}\ \emph {et~al.}(2023{\natexlab{c}})\citenamefont {Zhang}, \citenamefont {Lin}, \citenamefont {Moreo}, \citenamefont {Maier},\ and\ \citenamefont {Dagotto}}]{zhang2023structural}%
  \BibitemOpen
  \bibfield  {author} {\bibinfo {author} {\bibfnamefont {Y.}~\bibnamefont {Zhang}}, \bibinfo {author} {\bibfnamefont {L.-F.}\ \bibnamefont {Lin}}, \bibinfo {author} {\bibfnamefont {A.}~\bibnamefont {Moreo}}, \bibinfo {author} {\bibfnamefont {T.~A.}\ \bibnamefont {Maier}}, \ and\ \bibinfo {author} {\bibfnamefont {E.}~\bibnamefont {Dagotto}},\ }\href {https://arxiv.org/abs/2307.15276} {\bibfield  {journal} {\bibinfo  {journal} {arXiv:2307.15276}\ } (\bibinfo {year} {2023}{\natexlab{c}})}\BibitemShut {NoStop}%
\bibitem [{\citenamefont {Qu}\ \emph {et~al.}(2024)\citenamefont {Qu}, \citenamefont {Qu}, \citenamefont {Chen}, \citenamefont {Wu}, \citenamefont {Yang}, \citenamefont {Li},\ and\ \citenamefont {Su}}]{qu2023bilayer}%
  \BibitemOpen
  \bibfield  {author} {\bibinfo {author} {\bibfnamefont {X.-Z.}\ \bibnamefont {Qu}}, \bibinfo {author} {\bibfnamefont {D.-W.}\ \bibnamefont {Qu}}, \bibinfo {author} {\bibfnamefont {J.}~\bibnamefont {Chen}}, \bibinfo {author} {\bibfnamefont {C.}~\bibnamefont {Wu}}, \bibinfo {author} {\bibfnamefont {F.}~\bibnamefont {Yang}}, \bibinfo {author} {\bibfnamefont {W.}~\bibnamefont {Li}}, \ and\ \bibinfo {author} {\bibfnamefont {G.}~\bibnamefont {Su}},\ }\href {\doibase 10.1103/PhysRevLett.132.036502} {\bibfield  {journal} {\bibinfo  {journal} {Phys. Rev. Lett.}\ }\textbf {\bibinfo {volume} {132}},\ \bibinfo {pages} {036502} (\bibinfo {year} {2024})}\BibitemShut {NoStop}%
\bibitem [{\citenamefont {Yang}\ \emph {et~al.}(2023{\natexlab{b}})\citenamefont {Yang}, \citenamefont {Zhang},\ and\ \citenamefont {Zhang}}]{yang2023minimal}%
  \BibitemOpen
  \bibfield  {author} {\bibinfo {author} {\bibfnamefont {Y.-F.}\ \bibnamefont {Yang}}, \bibinfo {author} {\bibfnamefont {G.-M.}\ \bibnamefont {Zhang}}, \ and\ \bibinfo {author} {\bibfnamefont {F.-C.}\ \bibnamefont {Zhang}},\ }\href {\doibase 10.1103/PhysRevB.108.L201108} {\bibfield  {journal} {\bibinfo  {journal} {Phys. Rev. B}\ }\textbf {\bibinfo {volume} {108}},\ \bibinfo {pages} {L201108} (\bibinfo {year} {2023}{\natexlab{b}})}\BibitemShut {NoStop}%
\bibitem [{\citenamefont {Zhang}\ \emph {et~al.}(2023{\natexlab{d}})\citenamefont {Zhang}, \citenamefont {Lin}, \citenamefont {Moreo}, \citenamefont {Maier},\ and\ \citenamefont {Dagotto}}]{zhang2023trends}%
  \BibitemOpen
  \bibfield  {author} {\bibinfo {author} {\bibfnamefont {Y.}~\bibnamefont {Zhang}}, \bibinfo {author} {\bibfnamefont {L.-F.}\ \bibnamefont {Lin}}, \bibinfo {author} {\bibfnamefont {A.}~\bibnamefont {Moreo}}, \bibinfo {author} {\bibfnamefont {T.~A.}\ \bibnamefont {Maier}}, \ and\ \bibinfo {author} {\bibfnamefont {E.}~\bibnamefont {Dagotto}},\ }\href {\doibase 10.1103/PhysRevB.108.165141} {\bibfield  {journal} {\bibinfo  {journal} {Phys. Rev. B}\ }\textbf {\bibinfo {volume} {108}},\ \bibinfo {pages} {165141} (\bibinfo {year} {2023}{\natexlab{d}})}\BibitemShut {NoStop}%
\bibitem [{\citenamefont {Lu}\ \emph {et~al.}(2023{\natexlab{a}})\citenamefont {Lu}, \citenamefont {Li}, \citenamefont {Zeng}, \citenamefont {Hou}, \citenamefont {Wang}, \citenamefont {Yang},\ and\ \citenamefont {You}}]{lu2023superconductivity}%
  \BibitemOpen
  \bibfield  {author} {\bibinfo {author} {\bibfnamefont {D.-C.}\ \bibnamefont {Lu}}, \bibinfo {author} {\bibfnamefont {M.}~\bibnamefont {Li}}, \bibinfo {author} {\bibfnamefont {Z.-Y.}\ \bibnamefont {Zeng}}, \bibinfo {author} {\bibfnamefont {W.}~\bibnamefont {Hou}}, \bibinfo {author} {\bibfnamefont {J.}~\bibnamefont {Wang}}, \bibinfo {author} {\bibfnamefont {F.}~\bibnamefont {Yang}}, \ and\ \bibinfo {author} {\bibfnamefont {Y.-Z.}\ \bibnamefont {You}},\ }\href {https://arxiv.org/abs/2308.11195} {\bibfield  {journal} {\bibinfo  {journal} {arXiv:2308.11195}\ } (\bibinfo {year} {2023}{\natexlab{a}})}\BibitemShut {NoStop}%
\bibitem [{\citenamefont {Tian}\ \emph {et~al.}(2023)\citenamefont {Tian}, \citenamefont {Chen}, \citenamefont {Wang}, \citenamefont {He},\ and\ \citenamefont {Lu}}]{tian2023correlation}%
  \BibitemOpen
  \bibfield  {author} {\bibinfo {author} {\bibfnamefont {Y.-H.}\ \bibnamefont {Tian}}, \bibinfo {author} {\bibfnamefont {Y.}~\bibnamefont {Chen}}, \bibinfo {author} {\bibfnamefont {J.-M.}\ \bibnamefont {Wang}}, \bibinfo {author} {\bibfnamefont {R.-Q.}\ \bibnamefont {He}}, \ and\ \bibinfo {author} {\bibfnamefont {Z.-Y.}\ \bibnamefont {Lu}},\ }\href {https://arxiv.org/abs/2308.09698} {\bibfield  {journal} {\bibinfo  {journal} {arXiv:2308.09698}\ } (\bibinfo {year} {2023})}\BibitemShut {NoStop}%
\bibitem [{\citenamefont {Huang}\ \emph {et~al.}(2023)\citenamefont {Huang}, \citenamefont {Wang},\ and\ \citenamefont {Zhou}}]{vortex2023huang}%
  \BibitemOpen
  \bibfield  {author} {\bibinfo {author} {\bibfnamefont {J.}~\bibnamefont {Huang}}, \bibinfo {author} {\bibfnamefont {Z.~D.}\ \bibnamefont {Wang}}, \ and\ \bibinfo {author} {\bibfnamefont {T.}~\bibnamefont {Zhou}},\ }\href {\doibase 10.1103/PhysRevB.108.174501} {\bibfield  {journal} {\bibinfo  {journal} {Phys. Rev. B}\ }\textbf {\bibinfo {volume} {108}},\ \bibinfo {pages} {174501} (\bibinfo {year} {2023})}\BibitemShut {NoStop}%
\bibitem [{\citenamefont {Jiang}\ \emph {et~al.}(2023)\citenamefont {Jiang}, \citenamefont {Jinning~Hou}, \citenamefont {Lang},\ and\ \citenamefont {Ku}}]{jiang2023screening}%
  \BibitemOpen
  \bibfield  {author} {\bibinfo {author} {\bibfnamefont {R.}~\bibnamefont {Jiang}}, \bibinfo {author} {\bibfnamefont {Z.~F.}\ \bibnamefont {Jinning~Hou}}, \bibinfo {author} {\bibfnamefont {Z.-J.}\ \bibnamefont {Lang}}, \ and\ \bibinfo {author} {\bibfnamefont {W.}~\bibnamefont {Ku}},\ }\href {https://arxiv.org/abs/2308.11614} {\bibfield  {journal} {\bibinfo  {journal} {arXiv:2308.11614}\ } (\bibinfo {year} {2023})}\BibitemShut {NoStop}%
\bibitem [{\citenamefont {Liao}\ \emph {et~al.}(2023)\citenamefont {Liao}, \citenamefont {Chen}, \citenamefont {Duan}, \citenamefont {Wang}, \citenamefont {Liu}, \citenamefont {Yu},\ and\ \citenamefont {Si}}]{liao2023correlations}%
  \BibitemOpen
  \bibfield  {author} {\bibinfo {author} {\bibfnamefont {Z.}~\bibnamefont {Liao}}, \bibinfo {author} {\bibfnamefont {L.}~\bibnamefont {Chen}}, \bibinfo {author} {\bibfnamefont {G.}~\bibnamefont {Duan}}, \bibinfo {author} {\bibfnamefont {Y.}~\bibnamefont {Wang}}, \bibinfo {author} {\bibfnamefont {C.}~\bibnamefont {Liu}}, \bibinfo {author} {\bibfnamefont {R.}~\bibnamefont {Yu}}, \ and\ \bibinfo {author} {\bibfnamefont {Q.}~\bibnamefont {Si}},\ }\href {\doibase 10.1103/PhysRevB.108.214522} {\bibfield  {journal} {\bibinfo  {journal} {Phys. Rev. B}\ }\textbf {\bibinfo {volume} {108}},\ \bibinfo {pages} {214522} (\bibinfo {year} {2023})}\BibitemShut {NoStop}%
\bibitem [{\citenamefont {Lu}\ \emph {et~al.}(2023{\natexlab{b}})\citenamefont {Lu}, \citenamefont {Pan}, \citenamefont {Yang},\ and\ \citenamefont {Wu}}]{liao2023interlayer}%
  \BibitemOpen
  \bibfield  {author} {\bibinfo {author} {\bibfnamefont {C.}~\bibnamefont {Lu}}, \bibinfo {author} {\bibfnamefont {Z.}~\bibnamefont {Pan}}, \bibinfo {author} {\bibfnamefont {F.}~\bibnamefont {Yang}}, \ and\ \bibinfo {author} {\bibfnamefont {C.}~\bibnamefont {Wu}},\ }\href {https://arxiv.org/abs/2307.14965} {\bibfield  {journal} {\bibinfo  {journal} {arXiv:2307.14965}\ } (\bibinfo {year} {2023}{\natexlab{b}})}\BibitemShut {NoStop}%
\bibitem [{\citenamefont {Oh}\ and\ \citenamefont {Zhang}(2023)}]{oh2023tj}%
  \BibitemOpen
  \bibfield  {author} {\bibinfo {author} {\bibfnamefont {H.}~\bibnamefont {Oh}}\ and\ \bibinfo {author} {\bibfnamefont {Y.-H.}\ \bibnamefont {Zhang}},\ }\href {\doibase 10.1103/PhysRevB.108.174511} {\bibfield  {journal} {\bibinfo  {journal} {Phys. Rev. B}\ }\textbf {\bibinfo {volume} {108}},\ \bibinfo {pages} {174511} (\bibinfo {year} {2023})}\BibitemShut {NoStop}%
\bibitem [{\citenamefont {Qin}\ and\ \citenamefont {Yang}(2023)}]{qin2023singlets}%
  \BibitemOpen
  \bibfield  {author} {\bibinfo {author} {\bibfnamefont {Q.}~\bibnamefont {Qin}}\ and\ \bibinfo {author} {\bibfnamefont {Y.-f.}\ \bibnamefont {Yang}},\ }\href {\doibase 10.1103/PhysRevB.108.L140504} {\bibfield  {journal} {\bibinfo  {journal} {Phys. Rev. B}\ }\textbf {\bibinfo {volume} {108}},\ \bibinfo {pages} {L140504} (\bibinfo {year} {2023})}\BibitemShut {NoStop}%
\bibitem [{\citenamefont {Sakakibara}\ \emph {et~al.}(2023)\citenamefont {Sakakibara}, \citenamefont {Kitamine}, \citenamefont {Ochi},\ and\ \citenamefont {Kuroki}}]{sakakibara2023hubbard}%
  \BibitemOpen
  \bibfield  {author} {\bibinfo {author} {\bibfnamefont {H.}~\bibnamefont {Sakakibara}}, \bibinfo {author} {\bibfnamefont {N.}~\bibnamefont {Kitamine}}, \bibinfo {author} {\bibfnamefont {M.}~\bibnamefont {Ochi}}, \ and\ \bibinfo {author} {\bibfnamefont {K.}~\bibnamefont {Kuroki}},\ }\href {https://arxiv.org/abs/2306.06039} {\bibfield  {journal} {\bibinfo  {journal} {arXiv:2306.06039}\ } (\bibinfo {year} {2023})}\BibitemShut {NoStop}%
\bibitem [{\citenamefont {Shen}\ \emph {et~al.}(2023{\natexlab{b}})\citenamefont {Shen}, \citenamefont {Qin},\ and\ \citenamefont {Zhang}}]{Yang2023dmrg}%
  \BibitemOpen
  \bibfield  {author} {\bibinfo {author} {\bibfnamefont {Y.}~\bibnamefont {Shen}}, \bibinfo {author} {\bibfnamefont {M.}~\bibnamefont {Qin}}, \ and\ \bibinfo {author} {\bibfnamefont {G.-M.}\ \bibnamefont {Zhang}},\ }\href {\doibase 10.1088/0256-307X/40/12/127401} {\bibfield  {journal} {\bibinfo  {journal} {Chin. Phys. Lett.}\ }\textbf {\bibinfo {volume} {40}},\ \bibinfo {eid} {127401} (\bibinfo {year} {2023}{\natexlab{b}})}\BibitemShut {NoStop}%
\bibitem [{\citenamefont {Shilenko}\ and\ \citenamefont {Leonov}(2023)}]{shilenko2023}%
  \BibitemOpen
  \bibfield  {author} {\bibinfo {author} {\bibfnamefont {D.~A.}\ \bibnamefont {Shilenko}}\ and\ \bibinfo {author} {\bibfnamefont {I.~V.}\ \bibnamefont {Leonov}},\ }\href {\doibase 10.1103/PhysRevB.108.125105} {\bibfield  {journal} {\bibinfo  {journal} {Phys. Rev. B}\ }\textbf {\bibinfo {volume} {108}},\ \bibinfo {pages} {125105} (\bibinfo {year} {2023})}\BibitemShut {NoStop}%
\bibitem [{\citenamefont {Li}\ \emph {et~al.}(2024)\citenamefont {Li}, \citenamefont {Zhang}, \citenamefont {Xiang}, \citenamefont {Zhang}, \citenamefont {Zhu},\ and\ \citenamefont {Wen}}]{li2023signature}%
  \BibitemOpen
  \bibfield  {author} {\bibinfo {author} {\bibfnamefont {Q.}~\bibnamefont {Li}}, \bibinfo {author} {\bibfnamefont {Y.-J.}\ \bibnamefont {Zhang}}, \bibinfo {author} {\bibfnamefont {Z.-N.}\ \bibnamefont {Xiang}}, \bibinfo {author} {\bibfnamefont {Y.}~\bibnamefont {Zhang}}, \bibinfo {author} {\bibfnamefont {X.}~\bibnamefont {Zhu}}, \ and\ \bibinfo {author} {\bibfnamefont {H.-H.}\ \bibnamefont {Wen}},\ }\href {\doibase 10.1088/0256-307X/41/1/017401} {\bibfield  {journal} {\bibinfo  {journal} {Chin. Phys. Lett.}\ }\textbf {\bibinfo {volume} {41}},\ \bibinfo {pages} {017401} (\bibinfo {year} {2024})}\BibitemShut {NoStop}%
\bibitem [{\citenamefont {Zhang}\ \emph {et~al.}(2023{\natexlab{e}})\citenamefont {Zhang}, \citenamefont {Pei}, \citenamefont {Du}, \citenamefont {Cao}, \citenamefont {Wang}, \citenamefont {Wu}, \citenamefont {Li}, \citenamefont {Zhao}, \citenamefont {Li}, \citenamefont {Cao}, \citenamefont {Zhu}, \citenamefont {Zhang}, \citenamefont {Yu}, \citenamefont {Cheng}, \citenamefont {Zhao}, \citenamefont {Chen}, \citenamefont {Guo}, \citenamefont {Yang},\ and\ \citenamefont {Qi}}]{zhang2023superconductivity}%
  \BibitemOpen
  \bibfield  {author} {\bibinfo {author} {\bibfnamefont {M.}~\bibnamefont {Zhang}}, \bibinfo {author} {\bibfnamefont {C.}~\bibnamefont {Pei}}, \bibinfo {author} {\bibfnamefont {X.}~\bibnamefont {Du}}, \bibinfo {author} {\bibfnamefont {Y.}~\bibnamefont {Cao}}, \bibinfo {author} {\bibfnamefont {Q.}~\bibnamefont {Wang}}, \bibinfo {author} {\bibfnamefont {J.}~\bibnamefont {Wu}}, \bibinfo {author} {\bibfnamefont {Y.}~\bibnamefont {Li}}, \bibinfo {author} {\bibfnamefont {Y.}~\bibnamefont {Zhao}}, \bibinfo {author} {\bibfnamefont {C.}~\bibnamefont {Li}}, \bibinfo {author} {\bibfnamefont {W.}~\bibnamefont {Cao}}, \bibinfo {author} {\bibfnamefont {S.}~\bibnamefont {Zhu}}, \bibinfo {author} {\bibfnamefont {Q.}~\bibnamefont {Zhang}}, \bibinfo {author} {\bibfnamefont {N.}~\bibnamefont {Yu}}, \bibinfo {author} {\bibfnamefont {P.}~\bibnamefont {Cheng}}, \bibinfo {author} {\bibfnamefont {J.}~\bibnamefont {Zhao}}, \bibinfo {author} {\bibfnamefont {Y.}~\bibnamefont {Chen}}, \bibinfo {author} {\bibfnamefont {H.}~\bibnamefont
  {Guo}}, \bibinfo {author} {\bibfnamefont {L.}~\bibnamefont {Yang}}, \ and\ \bibinfo {author} {\bibfnamefont {Y.}~\bibnamefont {Qi}},\ }\href {https://arxiv.org/abs/2311.07423} {\bibfield  {journal} {\bibinfo  {journal} {arXiv:2311.07423}\ } (\bibinfo {year} {2023}{\natexlab{e}})}\BibitemShut {NoStop}%
\bibitem [{\citenamefont {Zhu}\ \emph {et~al.}(2024)\citenamefont {Zhu}, \citenamefont {Zhang}, \citenamefont {Pan}, \citenamefont {Chen}, \citenamefont {Peng}, \citenamefont {Chen}, \citenamefont {Ren}, \citenamefont {Liu}, \citenamefont {Li}, \citenamefont {Xing}, \citenamefont {Han}, \citenamefont {Wang}, \citenamefont {Jia}, \citenamefont {Wo}, \citenamefont {Gu}, \citenamefont {Gu}, \citenamefont {Ji}, \citenamefont {Wang}, \citenamefont {Gou}, \citenamefont {Shen}, \citenamefont {Ying}, \citenamefont {Chen}, \citenamefont {Yang}, \citenamefont {Zheng}, \citenamefont {Zeng}, \citenamefont {Guo},\ and\ \citenamefont {Zhao}}]{zhu2024superconductivity}%
  \BibitemOpen
  \bibfield  {author} {\bibinfo {author} {\bibfnamefont {Y.}~\bibnamefont {Zhu}}, \bibinfo {author} {\bibfnamefont {E.}~\bibnamefont {Zhang}}, \bibinfo {author} {\bibfnamefont {B.}~\bibnamefont {Pan}}, \bibinfo {author} {\bibfnamefont {X.}~\bibnamefont {Chen}}, \bibinfo {author} {\bibfnamefont {D.}~\bibnamefont {Peng}}, \bibinfo {author} {\bibfnamefont {L.}~\bibnamefont {Chen}}, \bibinfo {author} {\bibfnamefont {H.}~\bibnamefont {Ren}}, \bibinfo {author} {\bibfnamefont {F.}~\bibnamefont {Liu}}, \bibinfo {author} {\bibfnamefont {N.}~\bibnamefont {Li}}, \bibinfo {author} {\bibfnamefont {Z.}~\bibnamefont {Xing}}, \bibinfo {author} {\bibfnamefont {J.}~\bibnamefont {Han}}, \bibinfo {author} {\bibfnamefont {J.}~\bibnamefont {Wang}}, \bibinfo {author} {\bibfnamefont {D.}~\bibnamefont {Jia}}, \bibinfo {author} {\bibfnamefont {H.}~\bibnamefont {Wo}}, \bibinfo {author} {\bibfnamefont {Y.}~\bibnamefont {Gu}}, \bibinfo {author} {\bibfnamefont {Y.}~\bibnamefont {Gu}}, \bibinfo {author} {\bibfnamefont {L.}~\bibnamefont
  {Ji}}, \bibinfo {author} {\bibfnamefont {W.}~\bibnamefont {Wang}}, \bibinfo {author} {\bibfnamefont {H.}~\bibnamefont {Gou}}, \bibinfo {author} {\bibfnamefont {Y.}~\bibnamefont {Shen}}, \bibinfo {author} {\bibfnamefont {T.}~\bibnamefont {Ying}}, \bibinfo {author} {\bibfnamefont {X.}~\bibnamefont {Chen}}, \bibinfo {author} {\bibfnamefont {W.}~\bibnamefont {Yang}}, \bibinfo {author} {\bibfnamefont {C.}~\bibnamefont {Zheng}}, \bibinfo {author} {\bibfnamefont {Q.}~\bibnamefont {Zeng}}, \bibinfo {author} {\bibfnamefont {J.-G.}\ \bibnamefont {Guo}}, \ and\ \bibinfo {author} {\bibfnamefont {J.}~\bibnamefont {Zhao}},\ }\href {https://arxiv.org/abs/2311.07353} {\bibfield  {journal} {\bibinfo  {journal} {arXiv:2311.07353}\ } (\bibinfo {year} {2024})}\BibitemShut {NoStop}%
\bibitem [{\citenamefont {Zhang}\ \emph {et~al.}(2020{\natexlab{a}})\citenamefont {Zhang}, \citenamefont {Phelan}, \citenamefont {Botana}, \citenamefont {Chen}, \citenamefont {Zheng}, \citenamefont {Krogstad}, \citenamefont {Wang}, \citenamefont {Qiu}, \citenamefont {Rodriguez-Rivera}, \citenamefont {Osborn}, \citenamefont {Rosenkranz}, \citenamefont {Norman},\ and\ \citenamefont {Mitchell}}]{Zhang2020Intertwined}%
  \BibitemOpen
  \bibfield  {author} {\bibinfo {author} {\bibfnamefont {J.}~\bibnamefont {Zhang}}, \bibinfo {author} {\bibfnamefont {D.}~\bibnamefont {Phelan}}, \bibinfo {author} {\bibfnamefont {A.~S.}\ \bibnamefont {Botana}}, \bibinfo {author} {\bibfnamefont {Y.-S.}\ \bibnamefont {Chen}}, \bibinfo {author} {\bibfnamefont {H.}~\bibnamefont {Zheng}}, \bibinfo {author} {\bibfnamefont {M.}~\bibnamefont {Krogstad}}, \bibinfo {author} {\bibfnamefont {S.~G.}\ \bibnamefont {Wang}}, \bibinfo {author} {\bibfnamefont {Y.}~\bibnamefont {Qiu}}, \bibinfo {author} {\bibfnamefont {J.~A.}\ \bibnamefont {Rodriguez-Rivera}}, \bibinfo {author} {\bibfnamefont {R.}~\bibnamefont {Osborn}}, \bibinfo {author} {\bibfnamefont {S.}~\bibnamefont {Rosenkranz}}, \bibinfo {author} {\bibfnamefont {M.~R.}\ \bibnamefont {Norman}}, \ and\ \bibinfo {author} {\bibfnamefont {J.~F.}\ \bibnamefont {Mitchell}},\ }\href {\doibase 10.1038/s41467-020-19836-0} {\bibfield  {journal} {\bibinfo  {journal} {Nat. Commun}\ }\textbf {\bibinfo {volume} {11}},\ \bibinfo
  {pages} {6003} (\bibinfo {year} {2020}{\natexlab{a}})}\BibitemShut {NoStop}%
\bibitem [{\citenamefont {LaBollita}\ \emph {et~al.}(2023)\citenamefont {LaBollita}, \citenamefont {Pardo}, \citenamefont {Norman},\ and\ \citenamefont {Botana}}]{labollita2023electronic}%
  \BibitemOpen
  \bibfield  {author} {\bibinfo {author} {\bibfnamefont {H.}~\bibnamefont {LaBollita}}, \bibinfo {author} {\bibfnamefont {V.}~\bibnamefont {Pardo}}, \bibinfo {author} {\bibfnamefont {M.~R.}\ \bibnamefont {Norman}}, \ and\ \bibinfo {author} {\bibfnamefont {A.~S.}\ \bibnamefont {Botana}},\ }\href {https://arxiv.org/abs/2309.17279} {\bibfield  {journal} {\bibinfo  {journal} {arXiv:2309.17279}\ } (\bibinfo {year} {2023})}\BibitemShut {NoStop}%
\bibitem [{\citenamefont {Mukuda}\ \emph {et~al.}(2012)\citenamefont {Mukuda}, \citenamefont {Shimizu}, \citenamefont {Iyo},\ and\ \citenamefont {Kitaoka}}]{multilayer_cuprates}%
  \BibitemOpen
  \bibfield  {author} {\bibinfo {author} {\bibfnamefont {H.}~\bibnamefont {Mukuda}}, \bibinfo {author} {\bibfnamefont {S.}~\bibnamefont {Shimizu}}, \bibinfo {author} {\bibfnamefont {A.}~\bibnamefont {Iyo}}, \ and\ \bibinfo {author} {\bibfnamefont {Y.}~\bibnamefont {Kitaoka}},\ }\href {\doibase 10.1143/JPSJ.81.011008} {\bibfield  {journal} {\bibinfo  {journal} {J. Phys. Soc. Japan}\ }\textbf {\bibinfo {volume} {81}},\ \bibinfo {pages} {011008} (\bibinfo {year} {2012})}\BibitemShut {NoStop}%
\bibitem [{\citenamefont {Hohenberg}\ and\ \citenamefont {Kohn}(1964)}]{hohenberg1964inhomogeneous}%
  \BibitemOpen
  \bibfield  {author} {\bibinfo {author} {\bibfnamefont {P.}~\bibnamefont {Hohenberg}}\ and\ \bibinfo {author} {\bibfnamefont {W.}~\bibnamefont {Kohn}},\ }\href {\doibase 10.1103/PhysRev.136.B864} {\bibfield  {journal} {\bibinfo  {journal} {Phys. Rev.}\ }\textbf {\bibinfo {volume} {136}},\ \bibinfo {pages} {B864} (\bibinfo {year} {1964})}\BibitemShut {NoStop}%
\bibitem [{\citenamefont {Kohn}\ and\ \citenamefont {Sham}(1965)}]{kohn1965self}%
  \BibitemOpen
  \bibfield  {author} {\bibinfo {author} {\bibfnamefont {W.}~\bibnamefont {Kohn}}\ and\ \bibinfo {author} {\bibfnamefont {L.~J.}\ \bibnamefont {Sham}},\ }\href {\doibase 10.1103/PhysRev.140.A1133} {\bibfield  {journal} {\bibinfo  {journal} {Phys. Rev.}\ }\textbf {\bibinfo {volume} {140}},\ \bibinfo {pages} {A1133} (\bibinfo {year} {1965})}\BibitemShut {NoStop}%
\bibitem [{\citenamefont {Zhang}\ \emph {et~al.}(2020{\natexlab{b}})\citenamefont {Zhang}, \citenamefont {Zheng}, \citenamefont {Chen}, \citenamefont {Ren}, \citenamefont {Yonemura}, \citenamefont {Huq},\ and\ \citenamefont {Mitchell}}]{Zhang2020oxygen}%
  \BibitemOpen
  \bibfield  {author} {\bibinfo {author} {\bibfnamefont {J.}~\bibnamefont {Zhang}}, \bibinfo {author} {\bibfnamefont {H.}~\bibnamefont {Zheng}}, \bibinfo {author} {\bibfnamefont {Y.-S.}\ \bibnamefont {Chen}}, \bibinfo {author} {\bibfnamefont {Y.}~\bibnamefont {Ren}}, \bibinfo {author} {\bibfnamefont {M.}~\bibnamefont {Yonemura}}, \bibinfo {author} {\bibfnamefont {A.}~\bibnamefont {Huq}}, \ and\ \bibinfo {author} {\bibfnamefont {J.~F.}\ \bibnamefont {Mitchell}},\ }\href {\doibase 10.1103/PhysRevMaterials.4.083402} {\bibfield  {journal} {\bibinfo  {journal} {Phys. Rev. Mater.}\ }\textbf {\bibinfo {volume} {4}},\ \bibinfo {pages} {083402} (\bibinfo {year} {2020}{\natexlab{b}})}\BibitemShut {NoStop}%
\bibitem [{\citenamefont {Jung}\ \emph {et~al.}(2022)\citenamefont {Jung}, \citenamefont {Kapeghian}, \citenamefont {Hanson}, \citenamefont {Pamuk},\ and\ \citenamefont {Botana}}]{jung2022RPs}%
  \BibitemOpen
  \bibfield  {author} {\bibinfo {author} {\bibfnamefont {M.-C.}\ \bibnamefont {Jung}}, \bibinfo {author} {\bibfnamefont {J.}~\bibnamefont {Kapeghian}}, \bibinfo {author} {\bibfnamefont {C.}~\bibnamefont {Hanson}}, \bibinfo {author} {\bibfnamefont {B.}~\bibnamefont {Pamuk}}, \ and\ \bibinfo {author} {\bibfnamefont {A.~S.}\ \bibnamefont {Botana}},\ }\href {\doibase 10.1103/PhysRevB.105.085150} {\bibfield  {journal} {\bibinfo  {journal} {Phys. Rev. B}\ }\textbf {\bibinfo {volume} {105}},\ \bibinfo {pages} {085150} (\bibinfo {year} {2022})}\BibitemShut {NoStop}%
\bibitem [{\citenamefont {Kresse}\ and\ \citenamefont {Hafner}(1993)}]{Kresse:1993bz}%
  \BibitemOpen
  \bibfield  {author} {\bibinfo {author} {\bibfnamefont {G.}~\bibnamefont {Kresse}}\ and\ \bibinfo {author} {\bibfnamefont {J.}~\bibnamefont {Hafner}},\ }\href {\doibase 10.1103/PhysRevB.47.558} {\bibfield  {journal} {\bibinfo  {journal} {Phys. Rev. B}\ }\textbf {\bibinfo {volume} {47}},\ \bibinfo {pages} {558} (\bibinfo {year} {1993})}\BibitemShut {NoStop}%
\bibitem [{\citenamefont {Kresse}\ and\ \citenamefont {Joubert}(1999)}]{Kresse:1999dk}%
  \BibitemOpen
  \bibfield  {author} {\bibinfo {author} {\bibfnamefont {G.}~\bibnamefont {Kresse}}\ and\ \bibinfo {author} {\bibfnamefont {D.}~\bibnamefont {Joubert}},\ }\href {\doibase 10.1103/PhysRevB.59.1758} {\bibfield  {journal} {\bibinfo  {journal} {Phys. Rev. B}\ }\textbf {\bibinfo {volume} {59}},\ \bibinfo {pages} {1758} (\bibinfo {year} {1999})}\BibitemShut {NoStop}%
\bibitem [{\citenamefont {Kresse}\ and\ \citenamefont {Furthm\"uller}(1996)}]{Kresse:1996kl}%
  \BibitemOpen
  \bibfield  {author} {\bibinfo {author} {\bibfnamefont {G.}~\bibnamefont {Kresse}}\ and\ \bibinfo {author} {\bibfnamefont {J.}~\bibnamefont {Furthm\"uller}},\ }\href {\doibase 10.1103/PhysRevB.54.11169} {\bibfield  {journal} {\bibinfo  {journal} {Phys. Rev. B}\ }\textbf {\bibinfo {volume} {54}},\ \bibinfo {pages} {11169} (\bibinfo {year} {1996})}\BibitemShut {NoStop}%
\bibitem [{\citenamefont {Perdew}\ \emph {et~al.}(1996)\citenamefont {Perdew}, \citenamefont {Burke},\ and\ \citenamefont {Ernzerhof}}]{gga_pbe}%
  \BibitemOpen
  \bibfield  {author} {\bibinfo {author} {\bibfnamefont {J.~P.}\ \bibnamefont {Perdew}}, \bibinfo {author} {\bibfnamefont {K.}~\bibnamefont {Burke}}, \ and\ \bibinfo {author} {\bibfnamefont {M.}~\bibnamefont {Ernzerhof}},\ }\href {\doibase 10.1103/PhysRevLett.77.3865} {\bibfield  {journal} {\bibinfo  {journal} {Phys. Rev. Lett.}\ }\textbf {\bibinfo {volume} {77}},\ \bibinfo {pages} {3865} (\bibinfo {year} {1996})}\BibitemShut {NoStop}%
\bibitem [{\citenamefont {Togo}\ and\ \citenamefont {Tanaka}(2015)}]{phonopy}%
  \BibitemOpen
  \bibfield  {author} {\bibinfo {author} {\bibfnamefont {A.}~\bibnamefont {Togo}}\ and\ \bibinfo {author} {\bibfnamefont {I.}~\bibnamefont {Tanaka}},\ }\href {\doibase https://doi.org/10.1016/j.scriptamat.2015.07.021} {\bibfield  {journal} {\bibinfo  {journal} {Scripta Materialia}\ }\textbf {\bibinfo {volume} {108}},\ \bibinfo {pages} {1} (\bibinfo {year} {2015})}\BibitemShut {NoStop}%
\bibitem [{\citenamefont {Blaha}\ \emph {et~al.}(2020)\citenamefont {Blaha}, \citenamefont {Schwarz}, \citenamefont {Tran}, \citenamefont {Laskowski}, \citenamefont {Madsen},\ and\ \citenamefont {Marks}}]{Blaha2020wien2k}%
  \BibitemOpen
  \bibfield  {author} {\bibinfo {author} {\bibfnamefont {P.}~\bibnamefont {Blaha}}, \bibinfo {author} {\bibfnamefont {K.}~\bibnamefont {Schwarz}}, \bibinfo {author} {\bibfnamefont {F.}~\bibnamefont {Tran}}, \bibinfo {author} {\bibfnamefont {R.}~\bibnamefont {Laskowski}}, \bibinfo {author} {\bibfnamefont {G.~K.~H.}\ \bibnamefont {Madsen}}, \ and\ \bibinfo {author} {\bibfnamefont {L.~D.}\ \bibnamefont {Marks}},\ }\href {\doibase 10.1063/1.5143061} {\bibfield  {journal} {\bibinfo  {journal} {J. Chem. Phys.}\ }\textbf {\bibinfo {volume} {152}},\ \bibinfo {pages} {074101} (\bibinfo {year} {2020})}\BibitemShut {NoStop}%
\bibitem [{\citenamefont {Czy\ifmmode~\dot{z}\else \.{z}\fi{}yk}\ and\ \citenamefont {Sawatzky}(1994)}]{ldau_amf}%
  \BibitemOpen
  \bibfield  {author} {\bibinfo {author} {\bibfnamefont {M.~T.}\ \bibnamefont {Czy\ifmmode~\dot{z}\else \.{z}\fi{}yk}}\ and\ \bibinfo {author} {\bibfnamefont {G.~A.}\ \bibnamefont {Sawatzky}},\ }\href {\doibase 10.1103/PhysRevB.49.14211} {\bibfield  {journal} {\bibinfo  {journal} {Phys. Rev. B}\ }\textbf {\bibinfo {volume} {49}},\ \bibinfo {pages} {14211} (\bibinfo {year} {1994})}\BibitemShut {NoStop}%
\bibitem [{\citenamefont {Orobengoa}\ \emph {et~al.}(2009)\citenamefont {Orobengoa}, \citenamefont {Capillas}, \citenamefont {Aroyo},\ and\ \citenamefont {Perez-Mato}}]{amplimodes1}%
  \BibitemOpen
  \bibfield  {author} {\bibinfo {author} {\bibfnamefont {D.}~\bibnamefont {Orobengoa}}, \bibinfo {author} {\bibfnamefont {C.}~\bibnamefont {Capillas}}, \bibinfo {author} {\bibfnamefont {M.~I.}\ \bibnamefont {Aroyo}}, \ and\ \bibinfo {author} {\bibfnamefont {J.~M.}\ \bibnamefont {Perez-Mato}},\ }\href {\doibase 10.1107/S0021889809028064} {\bibfield  {journal} {\bibinfo  {journal} {J. Appl. Crystallogr.}\ }\textbf {\bibinfo {volume} {42}},\ \bibinfo {pages} {820} (\bibinfo {year} {2009})}\BibitemShut {NoStop}%
\bibitem [{\citenamefont {Perez-Mato}\ \emph {et~al.}(2010)\citenamefont {Perez-Mato}, \citenamefont {Orobengoa},\ and\ \citenamefont {Aroyo}}]{amplimodes2}%
  \BibitemOpen
  \bibfield  {author} {\bibinfo {author} {\bibfnamefont {J.~M.}\ \bibnamefont {Perez-Mato}}, \bibinfo {author} {\bibfnamefont {D.}~\bibnamefont {Orobengoa}}, \ and\ \bibinfo {author} {\bibfnamefont {M.~I.}\ \bibnamefont {Aroyo}},\ }\href {\doibase 10.1107/S0108767310016247} {\bibfield  {journal} {\bibinfo  {journal} {Acta Cryst.}\ }\textbf {\bibinfo {volume} {66}},\ \bibinfo {pages} {558} (\bibinfo {year} {2010})}\BibitemShut {NoStop}%
\bibitem [{\citenamefont {Geisler}\ \emph {et~al.}(2023)\citenamefont {Geisler}, \citenamefont {Hamlin}, \citenamefont {Stewart}, \citenamefont {Hennig},\ and\ \citenamefont {Hirschfeld}}]{geisler2023structural}%
  \BibitemOpen
  \bibfield  {author} {\bibinfo {author} {\bibfnamefont {B.}~\bibnamefont {Geisler}}, \bibinfo {author} {\bibfnamefont {J.~J.}\ \bibnamefont {Hamlin}}, \bibinfo {author} {\bibfnamefont {G.~R.}\ \bibnamefont {Stewart}}, \bibinfo {author} {\bibfnamefont {R.~G.}\ \bibnamefont {Hennig}}, \ and\ \bibinfo {author} {\bibfnamefont {P.~J.}\ \bibnamefont {Hirschfeld}},\ }\href {https://arxiv.org/abs/2309.15078} {\bibfield  {journal} {\bibinfo  {journal} {arXiv:2309.15078}\ } (\bibinfo {year} {2023})}\BibitemShut {NoStop}%
\bibitem [{Note1()}]{Note1}%
  \BibitemOpen
  \bibinfo {note} {Further ambiguity has been introduced with the report of La$_{3}$Ni$_{2}$O$_{7}$ with a `1313' stacking sequence of single and trilayer nickel planes~\cite {chen2023polymorphism, puphal2023unconventional}. Here, we only study the bilayer nickelate in the conventional (`2222') stacking and leave the comparison to the alternate stacking for future work.}\BibitemShut {Stop}%
\bibitem [{\citenamefont {Li}\ \emph {et~al.}(2017)\citenamefont {Li}, \citenamefont {Zhou}, \citenamefont {Nummy}, \citenamefont {Zhang}, \citenamefont {Pardo}, \citenamefont {Pickett}, \citenamefont {Mitchell},\ and\ \citenamefont {Dessau}}]{Li2017Fermiology}%
  \BibitemOpen
  \bibfield  {author} {\bibinfo {author} {\bibfnamefont {H.}~\bibnamefont {Li}}, \bibinfo {author} {\bibfnamefont {X.}~\bibnamefont {Zhou}}, \bibinfo {author} {\bibfnamefont {T.}~\bibnamefont {Nummy}}, \bibinfo {author} {\bibfnamefont {J.}~\bibnamefont {Zhang}}, \bibinfo {author} {\bibfnamefont {V.}~\bibnamefont {Pardo}}, \bibinfo {author} {\bibfnamefont {W.~E.}\ \bibnamefont {Pickett}}, \bibinfo {author} {\bibfnamefont {J.~F.}\ \bibnamefont {Mitchell}}, \ and\ \bibinfo {author} {\bibfnamefont {D.~S.}\ \bibnamefont {Dessau}},\ }\href {\doibase 10.1038/s41467-017-00777-0} {\bibfield  {journal} {\bibinfo  {journal} {Nat. Commun}\ }\textbf {\bibinfo {volume} {8}},\ \bibinfo {pages} {704} (\bibinfo {year} {2017})}\BibitemShut {NoStop}%
\bibitem [{\citenamefont {Puggioni}\ and\ \citenamefont {Rondinelli}(2018)}]{Puggioni2018crystal}%
  \BibitemOpen
  \bibfield  {author} {\bibinfo {author} {\bibfnamefont {D.}~\bibnamefont {Puggioni}}\ and\ \bibinfo {author} {\bibfnamefont {J.~M.}\ \bibnamefont {Rondinelli}},\ }\href {\doibase 10.1103/PhysRevB.97.115116} {\bibfield  {journal} {\bibinfo  {journal} {Phys. Rev. B}\ }\textbf {\bibinfo {volume} {97}},\ \bibinfo {pages} {115116} (\bibinfo {year} {2018})}\BibitemShut {NoStop}%
\bibitem [{\citenamefont {Pardo}\ and\ \citenamefont {Pickett}(2010)}]{Pardo2010quantum}%
  \BibitemOpen
  \bibfield  {author} {\bibinfo {author} {\bibfnamefont {V.}~\bibnamefont {Pardo}}\ and\ \bibinfo {author} {\bibfnamefont {W.~E.}\ \bibnamefont {Pickett}},\ }\href {\doibase 10.1103/PhysRevLett.105.266402} {\bibfield  {journal} {\bibinfo  {journal} {Phys. Rev. Lett.}\ }\textbf {\bibinfo {volume} {105}},\ \bibinfo {pages} {266402} (\bibinfo {year} {2010})}\BibitemShut {NoStop}%
\bibitem [{\citenamefont {Leonov}(2024)}]{leonov2024electronic}%
  \BibitemOpen
  \bibfield  {author} {\bibinfo {author} {\bibfnamefont {I.~V.}\ \bibnamefont {Leonov}},\ }\href {https://arxiv.org/abs/2401.07350} {\bibfield  {journal} {\bibinfo  {journal} {arXiv:2401.07350}\ } (\bibinfo {year} {2024})}\BibitemShut {NoStop}%
\bibitem [{\citenamefont {Tian}\ \emph {et~al.}(2024)\citenamefont {Tian}, \citenamefont {Ma}, \citenamefont {Ming}, \citenamefont {Zheng},\ and\ \citenamefont {Li}}]{tian2024effective}%
  \BibitemOpen
  \bibfield  {author} {\bibinfo {author} {\bibfnamefont {P.-F.}\ \bibnamefont {Tian}}, \bibinfo {author} {\bibfnamefont {H.-T.}\ \bibnamefont {Ma}}, \bibinfo {author} {\bibfnamefont {X.}~\bibnamefont {Ming}}, \bibinfo {author} {\bibfnamefont {X.-J.}\ \bibnamefont {Zheng}}, \ and\ \bibinfo {author} {\bibfnamefont {H.}~\bibnamefont {Li}},\ }\href {https://arxiv.org/abs/2402.02351} {\bibfield  {journal} {\bibinfo  {journal} {arXiv:2402.02351}\ } (\bibinfo {year} {2024})}\BibitemShut {NoStop}%
\bibitem [{\citenamefont {Wang}\ \emph {et~al.}(2024)\citenamefont {Wang}, \citenamefont {Ouyang}, \citenamefont {He},\ and\ \citenamefont {Lu}}]{wang2024nonfermi}%
  \BibitemOpen
  \bibfield  {author} {\bibinfo {author} {\bibfnamefont {J.-X.}\ \bibnamefont {Wang}}, \bibinfo {author} {\bibfnamefont {Z.}~\bibnamefont {Ouyang}}, \bibinfo {author} {\bibfnamefont {R.-Q.}\ \bibnamefont {He}}, \ and\ \bibinfo {author} {\bibfnamefont {Z.-Y.}\ \bibnamefont {Lu}},\ }\href {https://arxiv.org/abs/2402.02581} {\bibfield  {journal} {\bibinfo  {journal} {arXiv:2402.02581}\ } (\bibinfo {year} {2024})}\BibitemShut {NoStop}%
\bibitem [{\citenamefont {Chen}\ \emph {et~al.}(2023{\natexlab{c}})\citenamefont {Chen}, \citenamefont {Zhang}, \citenamefont {Thind}, \citenamefont {Sharma}, \citenamefont {LaBollita}, \citenamefont {Peterson}, \citenamefont {Zheng}, \citenamefont {Phelan}, \citenamefont {Botana}, \citenamefont {Klie},\ and\ \citenamefont {Mitchell}}]{chen2023polymorphism}%
  \BibitemOpen
  \bibfield  {author} {\bibinfo {author} {\bibfnamefont {X.}~\bibnamefont {Chen}}, \bibinfo {author} {\bibfnamefont {J.}~\bibnamefont {Zhang}}, \bibinfo {author} {\bibfnamefont {A.~S.}\ \bibnamefont {Thind}}, \bibinfo {author} {\bibfnamefont {S.}~\bibnamefont {Sharma}}, \bibinfo {author} {\bibfnamefont {H.}~\bibnamefont {LaBollita}}, \bibinfo {author} {\bibfnamefont {G.}~\bibnamefont {Peterson}}, \bibinfo {author} {\bibfnamefont {H.}~\bibnamefont {Zheng}}, \bibinfo {author} {\bibfnamefont {D.}~\bibnamefont {Phelan}}, \bibinfo {author} {\bibfnamefont {A.~S.}\ \bibnamefont {Botana}}, \bibinfo {author} {\bibfnamefont {R.~F.}\ \bibnamefont {Klie}}, \ and\ \bibinfo {author} {\bibfnamefont {J.~F.}\ \bibnamefont {Mitchell}},\ }\href {https://doi.org/10.1021/jacs.3c14052} {\bibfield  {journal} {\bibinfo  {journal} {J. Am. Chem. Soc.}\ } (\bibinfo {year} {2023}{\natexlab{c}})}\BibitemShut {NoStop}%
\bibitem [{\citenamefont {Puphal}\ \emph {et~al.}(2023)\citenamefont {Puphal}, \citenamefont {Reiss}, \citenamefont {Enderlein}, \citenamefont {Wu}, \citenamefont {Khaliullin}, \citenamefont {Sundaramurthy}, \citenamefont {Priessnitz}, \citenamefont {Knauft}, \citenamefont {Richter}, \citenamefont {Isobe}, \citenamefont {van Aken}, \citenamefont {Takagi}, \citenamefont {Keimer}, \citenamefont {Suyolcu}, \citenamefont {Wehinger}, \citenamefont {Hansmann},\ and\ \citenamefont {Hepting}}]{puphal2023unconventional}%
  \BibitemOpen
  \bibfield  {author} {\bibinfo {author} {\bibfnamefont {P.}~\bibnamefont {Puphal}}, \bibinfo {author} {\bibfnamefont {P.}~\bibnamefont {Reiss}}, \bibinfo {author} {\bibfnamefont {N.}~\bibnamefont {Enderlein}}, \bibinfo {author} {\bibfnamefont {Y.-M.}\ \bibnamefont {Wu}}, \bibinfo {author} {\bibfnamefont {G.}~\bibnamefont {Khaliullin}}, \bibinfo {author} {\bibfnamefont {V.}~\bibnamefont {Sundaramurthy}}, \bibinfo {author} {\bibfnamefont {T.}~\bibnamefont {Priessnitz}}, \bibinfo {author} {\bibfnamefont {M.}~\bibnamefont {Knauft}}, \bibinfo {author} {\bibfnamefont {L.}~\bibnamefont {Richter}}, \bibinfo {author} {\bibfnamefont {M.}~\bibnamefont {Isobe}}, \bibinfo {author} {\bibfnamefont {P.~A.}\ \bibnamefont {van Aken}}, \bibinfo {author} {\bibfnamefont {H.}~\bibnamefont {Takagi}}, \bibinfo {author} {\bibfnamefont {B.}~\bibnamefont {Keimer}}, \bibinfo {author} {\bibfnamefont {Y.~E.}\ \bibnamefont {Suyolcu}}, \bibinfo {author} {\bibfnamefont {B.}~\bibnamefont {Wehinger}}, \bibinfo {author} {\bibfnamefont
  {P.}~\bibnamefont {Hansmann}}, \ and\ \bibinfo {author} {\bibfnamefont {M.}~\bibnamefont {Hepting}},\ }\href {https://arxiv.org/abs/2312.07341} {\bibfield  {journal} {\bibinfo  {journal} {arXiv:2312.07341}\ } (\bibinfo {year} {2023})}\BibitemShut {NoStop}%
\bibitem [{\citenamefont {Stokes}\ \emph {et~al.}(2006)\citenamefont {Stokes}, \citenamefont {Hatch}, \citenamefont {Campbell},\ and\ \citenamefont {Tanner}}]{isotropy}%
  \BibitemOpen
  \bibfield  {author} {\bibinfo {author} {\bibfnamefont {H.~T.}\ \bibnamefont {Stokes}}, \bibinfo {author} {\bibfnamefont {D.~M.}\ \bibnamefont {Hatch}}, \bibinfo {author} {\bibfnamefont {B.~J.}\ \bibnamefont {Campbell}}, \ and\ \bibinfo {author} {\bibfnamefont {D.~E.}\ \bibnamefont {Tanner}},\ }\href {\doibase https://doi.org/10.1107/S0021889806014075} {\bibfield  {journal} {\bibinfo  {journal} {J. Appl. Crystallogr.}\ }\textbf {\bibinfo {volume} {39}},\ \bibinfo {pages} {607} (\bibinfo {year} {2006})}\BibitemShut {NoStop}%
\end{thebibliography}%

\clearpage
\newpage

\onecolumngrid

\appendix

\section{\label{app:phonons_la4310}Lattice dynamics for La$_4$Ni$_3$O$_{10}$}
\begin{figure}[H]
    \centering
    \includegraphics[width=\columnwidth]{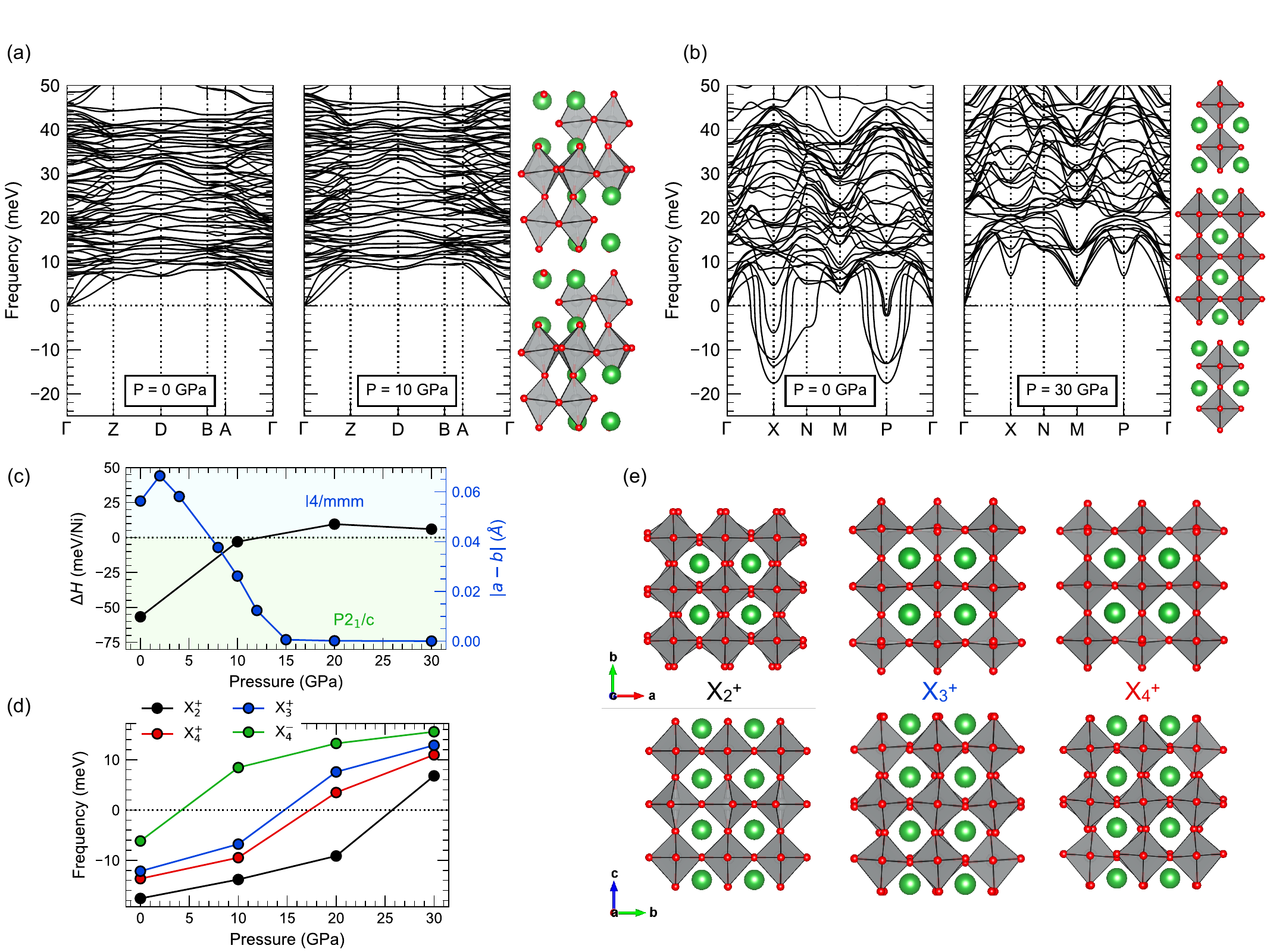}
    \caption{Lattice dynamics of La$_{4}$Ni$_{3}$O$_{10}$ in different crystal settings. Phonon dispersions for La$_{4}$Ni$_{3}$O$_{10}$ along high-symmetry lines for (a) monoclinic ($P2_{1}/c$) at ambient pressure (left) and 10 GPa (right) and (b) tetragonal ($I4/mmm$) at ambient (left) and 30 GPa (right) pressures. For the monoclinic structure, high-symmetry points in the Brillouin zone correspond to: $\Gamma$ = (0, 0, 0), Z = (0.0, 0.5, 0.0), D = (0.0, 0.5, 0.5), B = (0.0, 0.0, 0.5),  A = (-0.5, 0.0, 0.5). For the tetragonal structure, high-symmetry points in the Brillouin zone correspond to: $\Gamma$ = (0, 0, 0), X = (0.0, 0.0, 0.5), N = (0.0, 0.5, 0.0), M = (0.5, 0.5, -0.5), P = (0.25, 0.25, 0.25). (c) Enthalpy difference between the monoclinic and tetragonal structures and ``tetragonalization'' of the in-plane lattice parameters for the monoclinic structure as a function of pressure. (d) Phonons at the X point as a function of pressure. (e) Crystal distortions according to the X$_{2}^{+}$, X$_{3}^{+}$, and X$_{4}^{+}$ irreducible representations (irreps) of the $I4/mmm$ space group in the $ab$ (top) and $bc$ (bottom) planes.}
    \label{fig:4310-phonons-appendix}
\end{figure}

As described in Sec.~\ref{sec:structure}, the RP nickelates undergo a structural transition from a lower-symmetry phase to a higher-symmetry (tetragonal) phase upon hydrodstatic pressure. To understand this transition, we analyze the lattice dynamics of La$_{4}$Ni$_{3}$O$_{10}$ in both the low-symmetry (monoclinic) and high-symmetry (tetragonal) crystal settings. Fig.~\ref{fig:4310-phonons-appendix}(a) shows the phonon dispersion for the monoclinic structure at ambient and 10 GPa. No unstable modes are present in the phonon spectrum indicating the crystal structure is stable. In contrast, we compare these results with the phonon spectrum for the tetragonal structure with pressure in Fig.~\ref{fig:4310-phonons-appendix}(b). For the ambient pressure case, there are unstable modes at the X, N, and P points. In the main text, we focused on the X point as these modes correspond to the monoclinic distortions (see Fig.~\ref{fig:4310-phonons-appendix}(e)), while the N and P points do not based on a symmetry analysis using ISOTROPY~\cite{isotropy}. With hydrostatic pressure, the unstable modes become quenched (see Fig.~\ref{fig:4310-phonons-appendix}(d)) and we see that the $I4/mmm$ structure becomes energetically more favored based on enthalpy differences (see Fig.~\ref{fig:4310-phonons-appendix}(c)).

\section{\label{app:bands_la4310_space_group_dep}Band structures and constant-energy surfaces for La$_4$Ni$_3$O$_{10}$ in different space group symmetries}

\begin{figure}[H]
    \centering
    \includegraphics[width=\columnwidth]{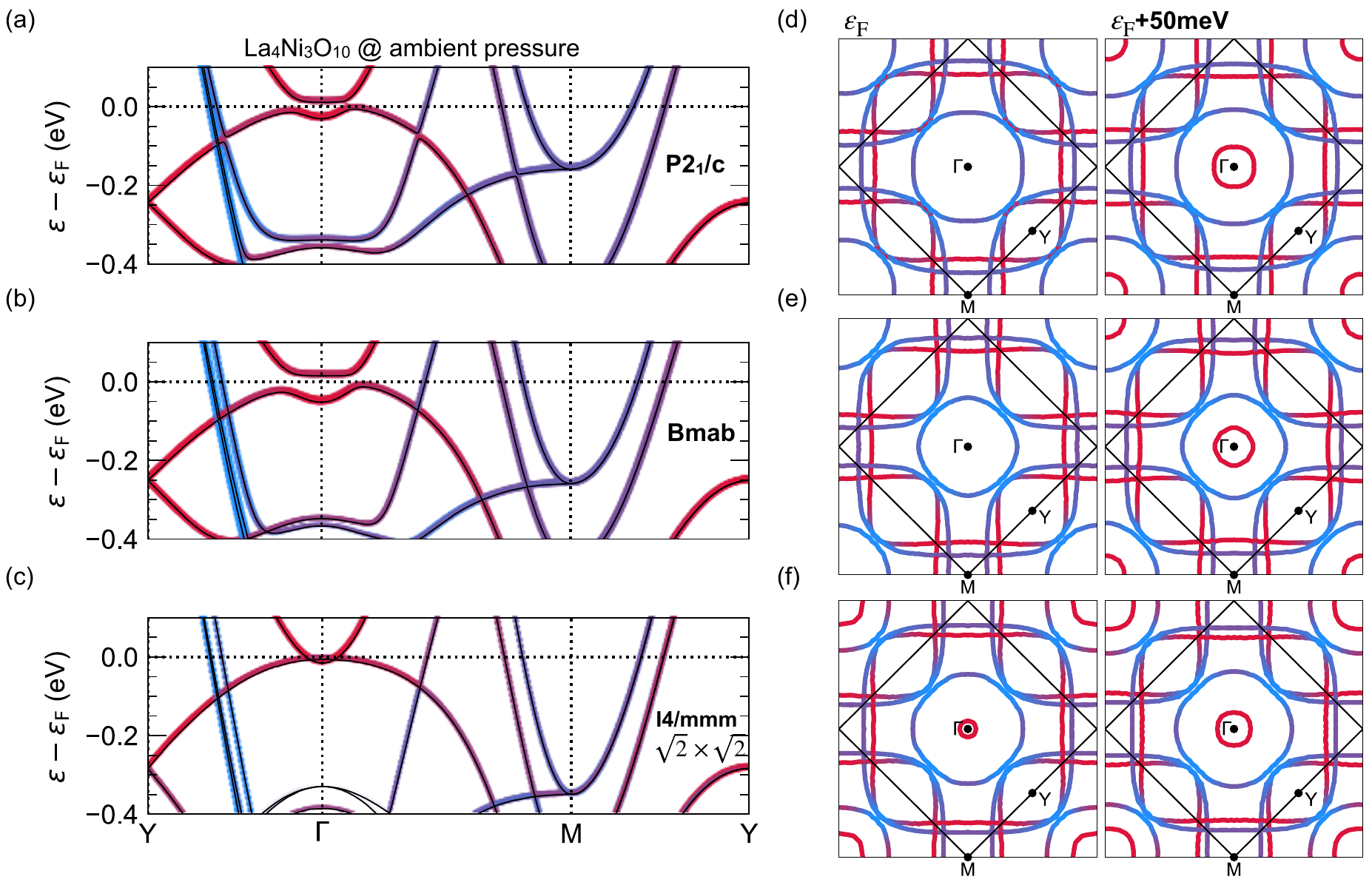}
    \caption{Low-energy electronic structure La$_{4}$Ni$_{3}$O$_{10}$ in different crystal settings at ambient pressure. Orbital-resolved (Ni-$e_{g}$ band structures in the $P2_{1}/c$ (a), $Bmab$ (b), and (c) $I4/mmm (\sqrt{2}\times\sqrt{2})$ crystal settings. Red (blue) corresponds to Ni-$d_{z^{2}}$ ($d_{x^{2}-y^{2}}$) orbital character. (d-f) Corresponding constant-energy surfaces in the $k_{z}=0$ plane at $\varepsilon_{\mathrm{F}}$ (left) and $\varepsilon_{\mathrm{F}}+50$ meV (right).}
    \label{fig:4310-YGMY}
\end{figure}

Fig.~\ref{fig:4310-YGMY} compares the computed band structure of La$_4$Ni$_3$O$_{10}$ at ambient pressure within different crystal settings. Importantly, the monoclinic ($P2_{1}/c$) and orthorhombic ($Bmab$) crystal phases exhibit the gapped $d_{z^{2}}$ band mentioned in the main text, while the tetragonal ($I4/mmm$) does not. This supports our argument that the gapping is due to band folding and is not associated with the charge density wave (CDW).

We note that in the lower-symmetry structures (monoclinic and orthorhombic), the gapping of the $d_{z^{2}}$ band removes the $\gamma$ pocket from the Fermi surface. This pocket is reintroduced in the tetragonal setting where this gap is closed (or with a small shift of the Fermi level (50 meV) as shown in Figs.~\ref{fig:4310-YGMY}(d-f)).

\section{\label{app:327}Electronic structure of \texorpdfstring{La$_{3}$Ni$_{2}$O$_{7}$}{La3Ni2O7}}
\begin{figure}
    \includegraphics[width=0.7\columnwidth]{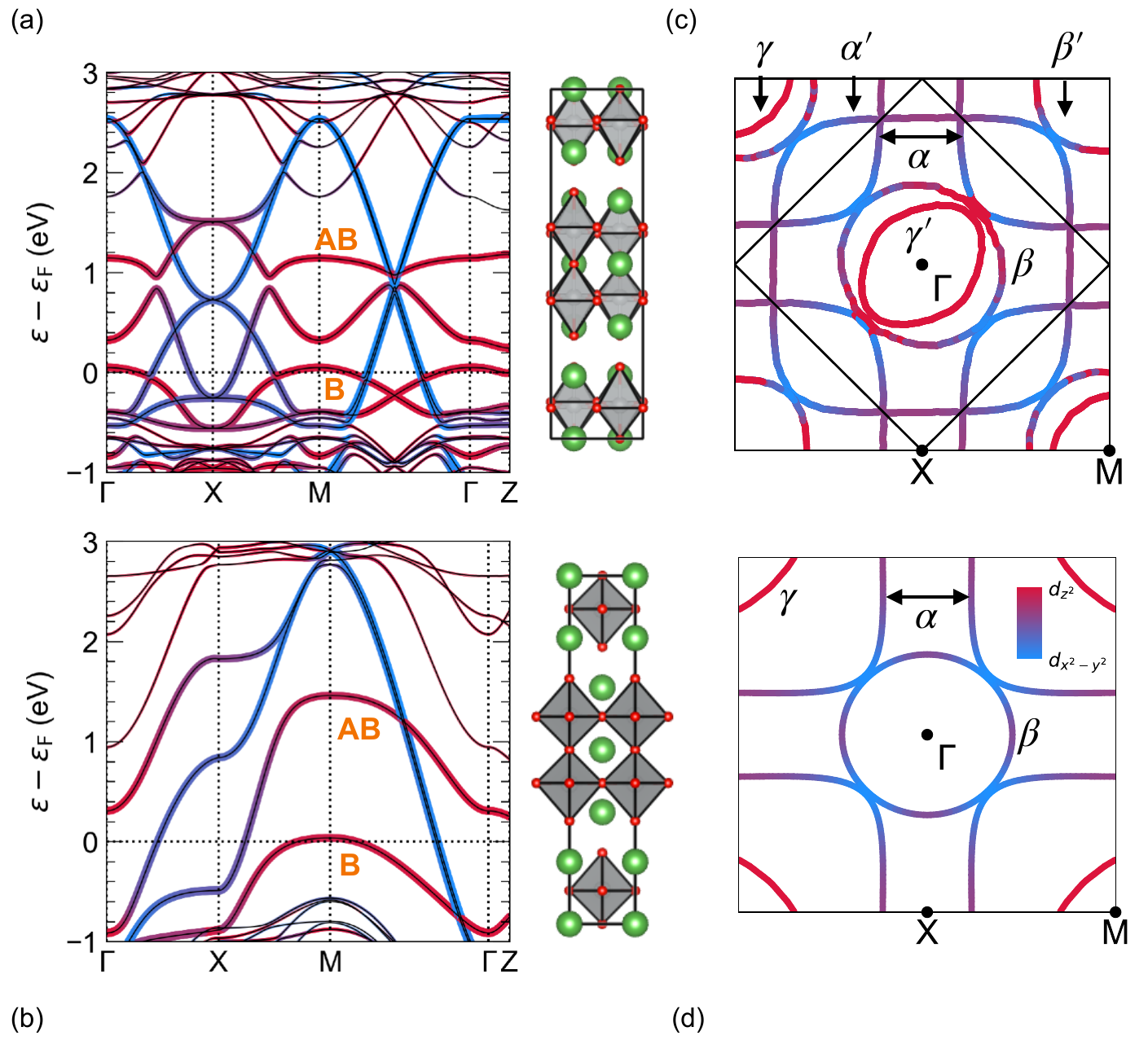}
    \caption{Electronic structure of La$_3$Ni$_2$O$_{7}$ at ambient pressure ($Amam$, top row) and $P=30$ GPa ($I4/mmm$, bottom row). (a,b) Orbital-resolved band structure along high-symmetry lines highlighting the Ni-$d_{z^{2}}$ (red) and Ni-$d_{x^{2}-y^{2}}$ (blue) orbital characters for the ambient ($Amam$) and 30 GPa ($I4/mmm$) phases, respectively. (c,d) Corresponding Fermi surfaces for (a,b) in the $k_{z}=0$ plane with high-symmetry points and $\alpha$, $\beta$, and $\gamma$ sheets denoted. Primed ($'$) sheets denote the backfolded version of the unprimed sheets for the $Amam$ case. Ni-$d_{z^{2}}$ bonding (B) and anti-bonding (AB) bands are labeled.}
    \label{fig:327-NM}
\end{figure}

Fig.~\ref{fig:327-NM} summarizes the paramagnetic electronic structure of La$_{3}$Ni$_{2}$O$_{7}$ at ambient pressure orthorhombic ($Amam$) and the high pressure (P = 30 GPa) tetragonal ($I4/mmm$) structures. 

As described before by us and others, focusing on the Ni-$e_{g}$ states near the Fermi level, the Ni-$d_{z^2}$ states are split by $\sim 1$ eV into an occupied bonding and unoccupied antibonding combination in the ambient pressure $Amam$ phase due to the quantum confinement provided by the nickel oxygen bilayers of the structure. 
The Ni-$d_{x^{2}-y^{2}}$ dispersion is large with a bandwidth of $\sim$2.5 eV and this orbital remains only partially occupied.
Turning to the electronic structure at high pressure (P = 30 GPa, $Fmmm$ phase), we find that the overall electronic structure within GGA is qualitatively similar to the ambient pressure {\it Amam} phase, with some quantitative differences. The Ni $d_{x^{2}-y^{2}}$ dispersion increases to 4 eV and the bonding-antibonding Ni-$d_{z^{2}}$ splitting increases to $\sim$ 1.5 eV. Similar to the ambient pressure case, the dominant density of states around the Fermi level for the nonmagnetic calculation at 29.5 GPa is that coming first from Ni-$d_{z^2}$ orbitals followed by that from the Ni-$d_{x^{2}-y^{2}}$ orbitals.

\section{\label{app:occupations}Occupations and magnetic moments in La$_4$Ni$_3$O$_{10}$}
Table~\ref{tab:on-site} summarizes layer-resolved Ni-$e_{g}$ moments for different magnetic configurations that are representative of the different spin states described in the purely ionic limit as shown in Fig.~\ref{fig:spin_states}. With La$_4$Ni$_3$O$_{10}$ being an itinerant system, there are some deviations with respect to the expected orbital occupation values in the pure ionic limit as shown in Fig.~\ref{fig:spin_states}. Interestingly, we note that the total Ni-$e_{g}$ occupation remains $\sim2$ which brings the filling of the Ni($3d$) shell to $\sim d^{8}$ which is the ionic configuration where one can differentiate HS and LS states. 

\begin{table}[H]
    \centering
    \begin{tabular*}{\columnwidth}
    {l@{\extracolsep{\fill}}ccccccc}
    \hline
    \hline
        solution (pressure) & spin state & layer &  $|m|$ ($\mu_{\mathrm{B}}$) & $d_{z^2}$ (up) & $d_{z^2}$ (down) & $d_{x^2-y^2}$ (up) & $d_{x^2-y^2}$ (down)\\
        \hline
        M/0/M  (ambient) & NM & inner  & 0.10 & 0.58 & 0.33 & 0.37 & 0.71\\
                        &  HS & outer  & 1.21 & 0.89 & 0.19 & 0.75 & 0.25\\
        FM (30 GPa)     &  HS & inner  & 1.09 & 0.77 & 0.22 & 0.74 & 0.20\\
                        &  HS & outer  & 1.23 & 0.87 & 0.19 & 0.75 & 0.23\\
        AFM-A (30 GPa)  &  HS & inner  & 0.97 & 0.79 & 0.29 & 0.64 & 0.22\\
                        &  HS & outer  & 1.12 & 0.86 & 0.24 & 0.71 & 0.21\\
        AFM-G (30 GPa)  &  LS/NM & inner  & 0.39 & 0.83 & 0.27 & 0.37 & 0.52\\
                        &  LS/NM & outer  & 0.39 & 0.84 & 0.25 & 0.36 & 0.57\\
        AFM-C (30 GPa)  &  LS & inner  & 0.64 & 0.58 & 0.33 & 0.72 & 0.32\\
                        &  LS & outer  & 0.91 & 0.84 & 0.22 & 0.65 & 0.36\\
    \hline
    \hline
    \end{tabular*}
    \caption{Layer-resolved Ni-$e_g$ occupations for the Ni atoms in different magnetic states for La$_4$Ni$_3$O$_{10}$ within GGA+$U$ ($U = 4$ eV, $J_{\mathrm{H}} = 0.7$ eV).}
    \label{tab:on-site}
\end{table}

\end{document}